\documentclass[journal]{IEEEtran}

\IEEEoverridecommandlockouts

\ifCLASSINFOpdf
\else
\fi

\usepackage{cite}
\usepackage{amsmath,amssymb,amsfonts}
\usepackage{algpseudocode}
\usepackage[linesnumbered,algoruled,boxed,lined]{algorithm2e}
\usepackage{graphicx}
\usepackage{textcomp}
\usepackage{xcolor}
\usepackage{multirow}
\usepackage{bm}
\usepackage{booktabs}
\usepackage{ntheorem}
\usepackage{ulem}
\usepackage{placeins}
\usepackage{tabularx}
\usepackage[caption=false]{subfig}

\theoremseparator{.}
\newskip\theorempreskipamount
\newskip\theorempostskipamount

\theorembodyfont{}

\allowdisplaybreaks[4]

\begin{document}
\pagestyle{empty} 

\title{Collaborative Charging Optimization for \\ Wireless Rechargeable Sensor Networks via Heterogeneous Mobile Chargers}

\author{Jianhang Yao, 
        Hui Kang, 
        Geng Sun, \IEEEmembership{Senior Member,~IEEE},
        Jiahui Li, \IEEEmembership{Member,~IEEE},\\
        Hongjuan Li, 
        Jiacheng Wang,
        Yinqiu Liu
\thanks{
        \par This work is supported in part by the National Key Research and Development Program of China under Grant 2024YFE03000104, in part by the Science and Technology Development Plan Project of Jilin Province under Grant 20250102210JC, in part by the Engineering Technology Research Center of Guangdong Provincial Key Scientific Research Platform Project under Grant 2024GCZX001, in part by Guangdong Provincial Key Disciplines Project under Grant 2024ZDJS142, in part by the Guangdong Provincial Higher Education Institutions Key Field Project on Next-generation Electronic Information (Semiconductor) under Grant 2025ZDZX1052, in part by Noncommunicable Chronic Diseases-National Science and Technology Major Project under Grant 2025ZD0551300, in part by Hainan S\&T Program under Grant ZDYF2025GXJS166, in part by Graduate Innovation Fund of Jilin University under Grant 2026CX212, and in part by the National Natural Science Foundation of China (62272194, 62471200).

        \par Jianhang Yao, Hui Kang, Jiahui Li, and Hongjuan Li are with the College of Computer Science and Technology, Jilin University, Changchun 130012, China (e-mails: yaojx25@mails.jlu.edu.cn; kanghui@jlu.edu.cn; lijiahui@jlu.edu.cn; hongjuan23@mails.jlu.edu.cn). 

        \par Geng Sun is with the College of Computer Science and Technology, Jilin University, Changchun 130012, China, and with Key Laboratory of Symbolic Computation and Knowledge Engineering of Ministry of Education, Jilin University, Changchun 130012, China (e-mail: sungeng@jlu.edu.cn).

        \par Jiacheng Wang and Yinqiu Liu are with the College of Computing and Data Science, Nanyang Technological University, Singapore 639798 (e-mails: jiacheng.wang@ntu.edu.sg; yinqiu001@ntu.edu.sg).
        % \par \textit{Corresponding authors: Geng Sun; Fang Mei} 
        \par Copyright (c) 2026 IEEE. Personal use of this material is permitted. However, permission to use this material for any other purposes must be obtained from the IEEE by sending a request to pubs-permissions@ieee.org.
        \par \textit{Corresponding authors: Geng Sun; Jiahui Li.} 
        \protect
	}

}

\IEEEtitleabstractindextext{%
\begin{abstract}
Despite the rapid proliferation of Internet of Things applications driving widespread wireless sensor network (WSN) deployment, traditional WSNs remain fundamentally constrained by persistent energy limitations that severely restrict network lifetime and operational sustainability. Wireless rechargeable sensor networks (WRSNs) integrated with wireless power transfer (WPT) technology emerge as a transformative paradigm, theoretically enabling unlimited operational lifetime. In this paper, we investigate a heterogeneous mobile charging architecture that strategically combines an automated aerial vehicle (AAV) and a ground smart vehicle (SV) in heterogeneous deployment scenarios to collaboratively exploit the superior mobility of the AAV and extended endurance of the SV for energy distribution. We formulate a multi-objective optimization problem that simultaneously addresses the dynamic balance of heterogeneous charger advantages, charging efficiency versus mobility energy consumption trade-offs, and real-time adaptive coordination under time-varying network conditions. This problem presents significant computational challenges due to its high-dimensional continuous action space, non-convex optimization landscape, and dynamic environmental constraints. To address these challenges, we propose the improved heterogeneous agent trust region policy optimization (IHATRPO) algorithm that integrates a self-attention mechanism for enhanced complex environmental state processing and employs a Beta sampling strategy to achieve unbiased gradient computation in continuous action spaces. Simulation results demonstrate that IHATRPO achieves a 51\% performance improvement over the original HATRPO, significantly outperforming state-of-the-art baseline algorithms while substantially decreasing sensor node mortality rate and improving charging system efficiency.

\end{abstract}


}

\maketitle

\IEEEdisplaynontitleabstractindextext
\IEEEpeerreviewmaketitle

% Section
% Introduction
%
\section{Introduction}

\par With the rapid proliferation of Internet of Things (IoT) applications, wireless sensor networks (WSNs) have become fundamental infrastructures for environmental monitoring, smart cities, industrial automation, and precision agriculture~\cite{Kandris2020, Greenwood2019}. WSNs are self-organizing wireless networks that monitor physical phenomena such as temperature, sound, vibration, or pollutants~\cite{Akyildiz2002}. Due to the small size, low power consumption, and autonomous network establishment capabilities of sensor nodes, conventional WSNs offer high flexibility, good adaptability, and low operational costs~\cite{Kandris2020}. However, WSNs face a persistent challenge as the finite energy capacity of sensor nodes severely constrains network lifetime and operational sustainability. Specifically, sensor nodes typically rely on batteries that are difficult or impossible to replace in remote deployments, thereby leading to network degradation and eventual failure as nodes exhaust their energy reserves~\cite{Li2022}. Recent research on extending the lifetime of WSNs has concentrated on energy conservation and energy provisioning approaches. While energy conservation techniques~\cite{Dhabliya2022, Khan2024} can significantly extend network lifetime, those methods cannot guarantee network stability since batteries will eventually be depleted. Energy provisioning through renewable energy harvesting offers a continuous energy supply, yet is constrained by unpredictable environmental conditions~\cite{Khan2024, LeonAvila2025}.

\par To address these fundamental limitations, wireless rechargeable sensor networks (WRSNs) have emerged as a transformative paradigm. Specifically, WRSNs integrate wireless power transfer (WPT) technology with conventional sensing capabilities, theoretically providing indefinite operational lifetime~\cite{Qureshi2022}. Moreover, WRSNs employ dedicated charging infrastructure that can be categorized into static charging stations and mobile charging platforms. Static charging stations, while providing reliable power delivery, require extensive deployment due to the limited spatial range of WPT technology, resulting in prohibitively high infrastructure costs and reduced deployment flexibility~\cite{Kaswan2022}. Conversely, mobile charging platforms offer superior coverage adaptability and dynamic resource allocation capabilities. Among mobile charging solutions, automated aerial vehicles (AAVs) and ground smart vehicles (SVs) represent two complementary approaches with distinct operational characteristics. Specifically, AAVs excel in mobility, rapid deployment, and terrain independence but are constrained by limited energy capacity and weather sensitivity~\cite{Sun2026, Mou2023}, while SVs provide extended operational endurance and robust performance but are restricted by terrain accessibility and mobility limitations~\cite{Lin2020}.

\par Current research on WRSNs focuses on single-type charging scenarios, which involve a trade-off between mobility and energy efficiency. However, single-type charging approaches, whether AAV-based or SV-based, cannot simultaneously optimize all critical performance metrics due to their individual limitations and the diverse requirements of WRSNs. Such fundamental limitations become particularly pronounced in complex deployment environments where sensor nodes exhibit varying energy demands, spatial distributions, and accessibility constraints that exceed the capabilities of any single charging platform. Motivated by these observations, we propose to combine AAV and SV platforms~\cite{Liu2022a} and design a heterogeneous mobile charging architecture to overcome the inherent limitations of homogeneous charging approaches. This strategic coordination between heterogeneous chargers enables adaptive resource allocation that responds to varying sensor node energy demands and environmental constraints, potentially revolutionizing the efficiency and reliability of WRSNs.

\par However, implementing such heterogeneous mobile charging coordination introduces several significant technical challenges that existing solutions cannot adequately address. \textit{Firstly}, the coordination problem between AAVs and SVs requires sophisticated collaborative decision-making mechanisms that can dynamically balance their respective advantages while accounting for different energy consumption patterns, mobility constraints, and charging capabilities in real-time operational conditions~\cite{Liu2022a}. \textit{Secondly}, the multi-objective optimization nature of the problem involves simultaneously maximizing charging efficiency, minimizing mobility energy consumption, and reducing sensor node mortality, then creating complex trade-offs that traditional optimization approaches cannot effectively resolve due to conflicting objectives and non-convex solution spaces~\cite{He2013}. \textit{Finally}, sensor network conditions exhibit dynamic and time-varying characteristics, including fluctuating energy levels, changing environmental conditions, and evolving communication requirements~\cite{Lin2016}, which necessitate adaptive strategies that can respond to these variations without compromising long-term performance objectives or system stability.

\par Accordingly, this paper proposes a novel deep reinforcement learning (DRL)-based approach for collaborative charging optimization in WRSNs employing heterogeneous mobile chargers. The main contributions of this paper are summarized as follows:

\begin{itemize}
    \item \textit{Innovative Heterogeneous Air-Ground Collaborative Charging System (HAGCCS) Model:} We design a comprehensive system model that strategically integrates the AAV and SV as collaborative charging agents in WRSNs. This architecture is specifically tailored for complex deployment scenarios where single-charger solutions prove inadequate. To the best of our knowledge, this is the first work to systematically investigate the collaborative charging optimization problem for heterogeneous mobile chargers while considering their distinctive mobility characteristics, energy constraints, and charging capabilities.
% R3-6

    \item \textit{Multi-Objective Optimization Problem with Heterogeneous Charger Interdependencies:} We formulate a multi-objective optimization problem that characterizes the complex interdependencies among charging efficiency maximization, mobility energy minimization, and sensor node mortality minimization in an environment with heterogeneous mobile chargers. This formulation enables the identification of fundamental trade-offs inherent in multi-objective optimization, where competing objectives generate a conflicting solution space, thus requiring collaborative coordination mechanisms. Moreover, this problem reveals distinctive coordination dynamics and complementary operational patterns in heterogeneous charger collaboration.

    \item \textit{DRL Solution with Heterogeneous Trust Region Strategy:} To address the dynamic and multi-objective nature of the optimization challenge, we propose the improved heterogeneous agent trust region policy optimization (IHATRPO) algorithm. This approach incorporates two key innovations. \textit{First}, the self-attention mechanism enables agents to process complex environmental information and inter-agent interactions more effectively. \textit{Second}, the Beta sampling strategy ensures unbiased gradient computation for continuous action spaces with bounded constraints. These enhancements specifically address the challenges of decentralized decision-making in heterogeneous multi-agent environments while ensuring convergence stability.

    \item \textit{Simulation and Performance Evaluation}: Simulation results demonstrate that the proposed algorithm outperforms various baselines, \textit{e.g.}, proximal policy optimization (PPO), multi-agent deep deterministic policy gradient (MADDPG), heterogeneous-agent trust region policy optimization (HATRPO). Moreover, the heterogeneous charger coordination approach significantly enhances sensor network survivability while maintaining charging efficiency. In addition, it is also confirmed that collaborative AAV-SV deployment provides adaptive coverage capabilities that effectively respond to dynamic network conditions.
\end{itemize}

\par The rest of this paper is organized as follows. Section \ref{sec:related_works} reviews the related research activities in WRSNs. Section \ref{sec:models_and_preliminaries} presents the system models. Section \ref{sec:problem_formulation_and_analysis} formulates the optimization problem. Section \ref{sec:solution} introduces the proposed IHATRPO algorithm. Section \ref{sec:simulation_results_and_analysis} provides the comprehensive simulation results and performance analysis, and Section \ref{sec:conclusion} concludes the paper with discussions on future research directions.

% Section
% Related Works
%
\section{Related Work}\label{sec:related_works}

\par In this work, we aim to propose a collaborative charging optimization framework in WRSNs by using heterogeneous mobile chargers. This topic involves the charging system architecture in WRSNs, optimization objectives in WRSN charging systems, and optimization methods for WRSN charging. Thus, we briefly introduce the related works of these areas as follows.

\subsection{Charging System Architectures in WRSNs}

\par Various charging system architectures have been designed to prolong the network lifetime in WRSNs. Traditional ground-based charging strategies have been extensively investigated, where mobile charging vehicles traverse the network to replenish sensor nodes. For example, the authors in~\cite{Dai2020} proposed a periodic charging and scheduling scheme aimed at optimizing the charging time and sensor selection of charging vehicles. Moreover, the authors in~\cite{Liang2021} proposed an on-demand charging strategy that incorporates spatial, temporal, and event domain characteristics of nodes, while utilizing an improved K-means algorithm for network partitioning with terrestrial wireless charging vehicles. Further building upon this ground-based mobile charger architecture, the authors in~\cite{Tao2020} focused on optimizing for network tasks by jointly selecting sensors and allocating energy.

\par With the advancement of AAV technology, aerial charging systems have emerged as promising alternatives for WRSN energy replenishment. For example, the authors in~\cite{Liu2022} proposed a joint scheduling and trajectory optimization problem for single-AAV based charging scenarios, thus improving charging efficiency by reducing repeated charging nodes while minimizing hovering points and flight distance. Furthermore, the authors in~\cite{Liang2021} investigated a multi-AAV deployment optimization problem and proposed an improved firefly algorithm to optimize charging efficiency, motion energy consumption, and sensor coverage. In~\cite{Ning2024}, the authors proposed a cooperative air-ground architecture where one AAV charges sensors, and a ground-based vehicle provides battery replacement for the AAV, using a Deep Q-Network to optimize the strategy.

\par Recent studies have further explored advanced scheduling under mobility constraints and probabilistic approaches. Specifically, the authors in~\cite{Qaisar2024a} proposed an integrated sensing and communication-assisted WRSN protocol that coordinates multiple mobile charging vehicles (MCVs) by incorporating probabilistic techniques to balance charging load and reduce travel cost. Moreover, the authors in~\cite{Qaisar2024} proposed a novel scheduling protocol, which models key node attributes as a probability distribution to guide on-demand MCV dispatch, improving charging delay and energy efficiency. Furthermore, the authors in~\cite{Zhao2024} investigated periodic charging scheduling in AAV-based WRSNs with automatic landing pads, extending AAV service range through mid-flight energy replenishment.

% R2-6

\par However, these works treat ground-based and aerial charging systems as independent solutions, thus overlooking the potential collaborative benefits of air-ground cooperative charging. Different from these methods, we design a heterogeneous charging system that simultaneously coordinates both the AAV and SV to achieve complementary operational advantages and compensate for individual limitations.

\subsection{Optimization Objectives in WRSN Charging Systems}

\par The optimization objectives in WRSN charging systems have been primarily focused on network lifetime maximization and node mortality rate minimization. For instance, the authors in~\cite{Yang2024} proposed a hybrid approach targeting network longevity through optimized charging scheduling, where inner rings adopt single-node charging with flat topology while outer rings employ multi-node charging with cluster topology. Moreover, the authors in~\cite{Lee2022} proposed an energy-efficient adaptive directional charging algorithm that focuses on maximizing sensor node survival rates by adaptively selecting single-node or multi-node charging based on sensor node density.

\par Energy consumption optimization of mobile chargers represents another critical research direction. The authors in~\cite{Zhang2024} proposed a DRL-based mobile safety policy intervention algorithm specifically targeting single mobile charger energy efficiency in an uncertain environment with mobile obstacles. Moreover, the authors in~\cite{Li2024} combined SV deployment with recovery operations, jointly optimizing charging and recovery scheduling to minimize overall system energy consumption while handling increased charging requests.

\par Charging efficiency has also received considerable attention in recent studies. Specifically, the authors in~\cite{Orumwense2022} proposed efficient algorithms for increasing energy efficiency in WRSNs for cyber-physical systems through intelligent scheduling and sensor node prioritization without requiring prior knowledge of energy levels. Furthermore, trajectory optimization has emerged as a key goal for enhancing charging efficiency, where researchers focus on minimizing travel distances and optimizing charging paths to improve overall system performance.

\par Recent multi-objective studies have further advanced AAV-assisted charging and IoT optimization. Specifically, the authors in~\cite{Yu2021} proposed a multi-objective DRL algorithm for an AAV-assisted wireless powered IoT network to jointly optimize data rate, harvested energy, and AAV energy consumption. Moreover, the authors in~\cite{Zhang2022} jointly optimized average data rate, energy consumption, and coverage fairness for AAV-assisted IoT networks via combined on-policy and off-policy RL. Furthermore, the authors in~\cite{Lyu2024} formulated a multi-objective resource allocation problem for an AAV-assisted power IoT system enabling simultaneous data collection and wireless charging, solved by an RL-based dynamic algorithm.

\par However, these studies either optimize a single charging-related objective or combine WRSN charging with other tasks. Different from these approaches, our work focuses on the charging tasks while jointly optimizing the multi-objective problem composed of the mortality of sensor nodes, energy consumption of chargers, and charging efficiency.

\subsection{Optimization Methods for WRSN Charging}

\par Conventional optimization methods have been widely employed for WRSN charging problems. For example, the graph-based optimization approaches have been extensively used, where the authors in~\cite{Lin2023a} proposed comprehensive frameworks by using hexagonal decomposition and boustrophedon path planning for energy-aware coordination of one AAV in WRSN, thus addressing simultaneous period-area coverage, charging scheduling, and resource allocation challenges. Moreover, evolutionary computation methods have also demonstrated effectiveness, as shown in~\cite{Li2022}, which proposed an improved non-dominated sorting genetic algorithm-based solution for many-objective charging optimization in WRSNs. Additionally, heuristic optimization techniques have been applied in some works, where researchers employ greedy algorithms and local search methods to solve charging scheduling problems with polynomial time complexity.

\par Recent advances in DRL have introduced intelligent decision-making capabilities to WRSN charging systems. For instance, the authors in~\cite{Jiang2024} proposed a novel DRL approach with a hybrid action space for mobile charging, specifically employing the deep deterministic policy gradient (DDPG) algorithm to determine optimal charging time allocation and achieve improved network lifetime through continuous action space control. Furthermore, the authors in~\cite{Liang2022} introduced an asynchronous and scalable multi-agent proximal policy optimization algorithm for cooperative charging, thus demonstrating enhanced charging coordination through distributed policy optimization with improved scalability for large-scale scenarios.

\par However, these DRL-based works primarily focus on homogeneous multi-agent systems without considering the coordination challenges inherent in heterogeneous agent environments. Current approaches lack the collaborative mechanisms required to handle heterogeneous agent coordination between the AAV and SV with fundamentally different operational characteristics. These limitations motivate us to propose a specialized multi-agent DRL algorithm capable of managing heterogeneous agent interactions.

\subsection{Motivation and Contributions of This Work}

\par Different from these works, we consider a heterogeneous air-ground cooperative charging system by using both the AAV and SV. Moreover, we formulate a multi-objective optimization problem that jointly considers the mortality of sensor nodes, energy consumption of chargers, and charging efficiency. To solve it, we propose an innovative heterogeneous multi-agent DRL method specifically designed for coordinating agents with diverse operational characteristics and capabilities. In the following section, therefore, we present a detailed description of the system model under consideration.

% Section
% System Models and Preliminaries
%

\section{System Models and Preliminaries} \label{sec:models_and_preliminaries}

% autonomous aerial vehicle
\par In this section, we introduce the models of the considered HAGCCS, including the network model, wireless charging model, and energy consumption models of the AAV and SV.
% R3-6

%
%Sensor Network Model
% Sensor Network Model -> Network Model
\subsection{Network Model}

\par
\begin{figure}
    \centering
    \includegraphics[width=0.8\linewidth]{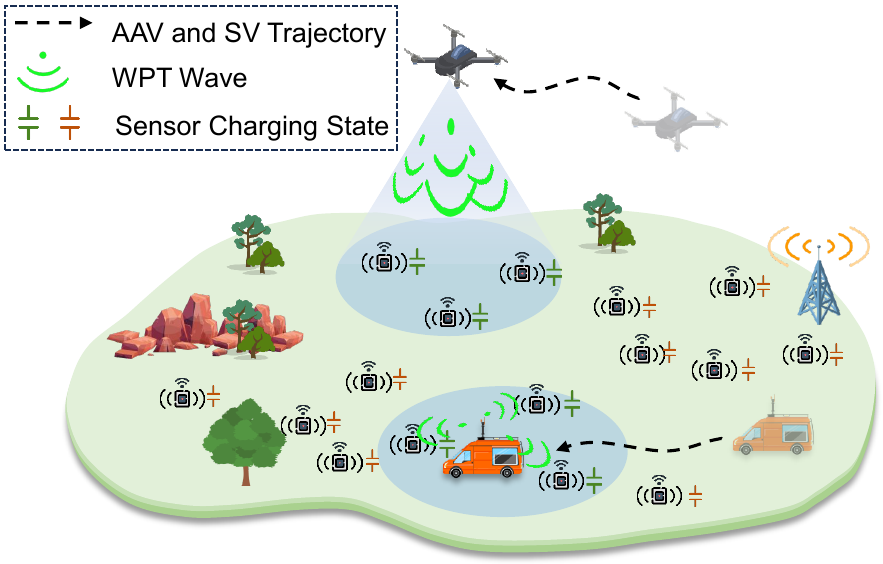}
    \caption{Architecture diagram of the HAGCCS for the WRSN. The AAV and SV travel within the WRSN to collaboratively provide energy for sensors through WPT waves.}
    \label{fig:model}
\end{figure}

 The HAGCCS under consideration is illustrated in Fig.~\ref{fig:model}, and it comprises the following elements:

\begin{itemize}

    \item A set of sensor nodes $\mathcal{S}=\{1,2,\dots,N_{\mathcal{S}}\}$. These sensor nodes are stationary and randomly distributed throughout the network, primarily tasked with data collection. Note that each sensor node can transmit data to a remote base station (BS) or receive commands from it~\cite{Liu2022a}. Moreover, each sensor node is equipped with an energy harvesting unit and an energy storage unit, which means that it can receive and store wireless energy transferred by mobile chargers~\cite{Ning2025}.
    % R3-6
    \item A pair of heterogeneous mobile chargers. Specifically, the heterogeneous mobile chargers consist of an AAV and an SV. Note that both the AAV and SV are capable of processing data from sensor nodes, BSs, and other mobile chargers~\cite{Chen2022}. Moreover, the AAV and SV can travel freely within the network area to provide charging service for the sensor nodes within a specified radius~\cite{Xie2012}, and their batteries power both of them. 
    % R3-6
    \item A remote BS that acts as a data fusion center. This BS is located at the edge of the region for data collection, and without loss of generality, we consider that the BS has no energy constraint since it has a sufficient energy supply~\cite{Yi2019}.
% R3-6

\end{itemize}

%
% table notation
%

\begin{table}[!h]
\centering
\renewcommand\arraystretch{1.15}
\caption{Summary of main notations}
\label{table:notation}
% \resizebox{3.5in}{!}{
\begin{tabular}{ll@{}}
\toprule
\textbf{Notation}           & \textbf{Description}                                 \\ \midrule
$\mu$                       & Charging efficiency        \\
$\lambda_1,\lambda_2,\lambda_3$ & Weighting coefficients for reward components \\
$\lambda$                   & Wavelength of RF \\
$\rho$                      & Air density   \\
$\eta$                      & Rectifier efficiency   \\
$\delta$                    & The KL threshold \\
$\theta_i$                  & Heading angle of agent $i$    \\
$\gamma$                    & Discount factor   \\
$\alpha_b,\beta_b$          & Shape parameters of Beta distribution        \\
$a_t^i$                     & Action of agent $i$ at time slot $t$ \\
$b_i$                       & Binary death indicator of sensor node $i$ \\
$d_{\max}$                  & Maximum charging radius of AAV/SV \\
$d_i$                       & Distance between sensor node and agent $i$      \\
$d$                         & Distance between sensor node and AAV/SV \\
% $\mathcal{D}$               & Trajectory buffer of AAV/SV \\
% $D_{KL}$                    & KL divergence        \\
$f_1, f_2, f_3$             & Charging efficiency, travel distance, node mortality\\
$G_s, G_r$                  & Antenna gain of transmitter and receiver\\
% $h_{AAV}$                         & Flight altitude of AAV    \\
% $k_1,k_2,k_3$               & Control parameters for SV motor   \\
$\mathcal{N}$               & Agent set        \\
$L_p$                       & Polarization loss \\
$P_0$                       & Transmit power of AAV/SV \\
$P_i$                       & Received power at sensor node $i$        \\
$P_{AAV}(v)$                & Motion energy consumption of AAV        \\
$P_{SV}(v)$                 & Motion energy consumption of SV        \\
$q_i^t$                     & Energy level of sensor node $i$ at time $t$   \\
$s_t$                       & State space at time slot $t$ \\
$\mathcal{S}$               & Set of sensor nodes        \\
$\mathcal{T}$               & Set of time slots        \\
$v$                         & Flight/travel speed of AAV/SV       \\
$X_{max},Y_{max}$           & Maximum range of WRSN area    \\
$\boldsymbol{Z}_t$          & Decision variables at time slot $t$  
 \\ \bottomrule
\end{tabular}
% }
\end{table}

\begin{figure}
    \centering
    \includegraphics[width=0.9\linewidth]{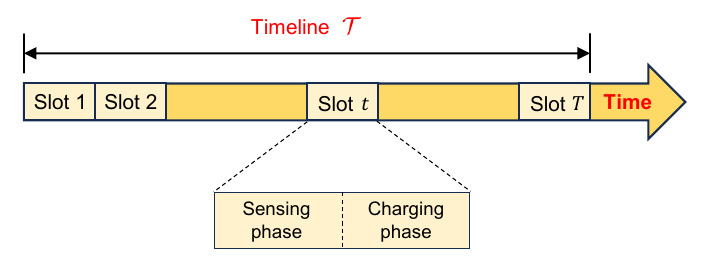}
    \caption{The time slot division model in HAGCCS.}
    \label{fig:timeline}
\end{figure}

\par In HAGCCS, the energy consumption of sensor nodes typically follows certain protocols and cycles to ensure efficient network operation and prolong network lifetime. In this case, we consider a discrete-time system evolving over the timeline $\mathcal{T} = \{t|1,2,..., T \}$. Specifically, each time slot $t$ consists of two main phases that are the sensing phase and charging phase, as illustrated in Fig.~\ref{fig:timeline}. In the sensing phase, sensor nodes perform data collection, data processing, and data transmission. In the charging phase, the AAV and SV provide wireless energy transfer to the sensor nodes.

\par Based on this, we consider that all the sensor nodes and SV are located within the same two-dimensional plane, while the AAV maintains a constant altitude when flying or hovering. As such, the locations of the $i$-th sensor node, AAV, SV, are denoted as $(x_{i}^{\mathcal{S}}, y_{i}^{\mathcal{S}}, 0)$, $(x^{AAV}, y^{AAV}, h)$, $(x^{SV}, y^{SV}, 0)$, respectively.

\par As such, during each time slot, the AAV and SV travel freely within the sensor network to charge nearby sensor nodes, which aims to improve the charging efficiency and extend the network lifetime. In the following, we model the wireless charging model and energy consumption model of the AAV and SV to identify the key decision variables for optimizing wireless energy transfer and its transmission efficiency.

%
% Wireless Charging Model
%
\subsection{Wireless Charging Model}\label{charging_model}

\par In WRSNs, WPT enables the transmission of electrical energy wirelessly from the transmitter to the receiver across the air gap. We consider a radio-frequency (RF) based omnidirectional WPT model~\cite{Mei2004}, which utilizes RF waves at a specific frequency for energy transmission, thereby allowing energy to propagate in all directions.

\par As such, the charging efficiency $\mu$ of the AAV or SV for sensor nodes can be defined as follows:

\begin{equation}\label{eq:charging_efficiency}
    \mu=\frac{G_sG_r\eta}{L_p} \left(\frac{\lambda}{4\pi(d+\beta)}\right)^2,
\end{equation}

\noindent where $G_s$ denotes the antenna gain of the AAV or SV, $G_r$ represents the antenna gain of the sensor nodes as the receiver, $\lambda$ is the wavelength of the RF signal, $\eta$ is the rectifier efficiency, $L_p$ is the polarization loss, $\beta$ is a tunable parameter in the Friis free-space equation, and $d$ is the distance between the AAV or SV and the sensor node. 

\par Since in Eq.~\eqref{eq:charging_efficiency}, all parameters except for $d$ and $\beta$ are constant values in a specific WRSN, the calculation for the charging efficiency $\mu$ can be simplified as $\mu=\alpha/(d+\beta)^2$, where $\alpha$ is a constant that encompasses the parameter values of $G_s$, $G_r$, $\eta$, $L_p$, $\lambda$, and others from Eq.~\eqref{eq:charging_efficiency}. Let $P_0$ represent the transmit power of the AAV or SV. Then, the received power $P_i$ at the $i$-th sensor node $S_i$ can be given by $P_i=\mu_iP_0$.

\par From Eq.~\eqref{eq:charging_efficiency}, it can be observed that the received power at the sensor node primarily depends on the distance between the AAV or SV and the sensor node, as all parameters except for $d$ can be considered constants. As such, we set the max charging distance $d_{max}$ to assess the impact of distance on the received power. Specifically, when the distance between the AAV (or SV) and sensor node exceeds $d_{max}$, the received power at the sensor node becomes too low for energy rectification, thus preventing effective charging. Therefore, $d_{max}$ can be regarded as the effective charging radius. The received power $P_i$ can then be further expressed as follows:

\begin{equation}\label{eq:charging_limitation}
    P_i=\begin{cases}\frac{\alpha P_0}{(d_i+\beta)^2}& d_i\leq d_{max}\\0& d_i>d_{max}\end{cases}.
\end{equation}

%
%Energy Consumption Model of AAV and SV
%
\subsection{Energy Consumption Model of AAV and SV}

\par The total energy consumption of the AAV and SV consists of two main components. The first part is the energy consumed by the AAV and SV for charging sensor nodes. The second part is the energy consumed during the movement of the AAV and SV, including propulsion and hovering for the AAV, as well as the travel of the SV. Moreover, the energy consumption caused by communication among sensor nodes, mobile chargers, and BS is negligible compared to the movement energy consumption. Therefore, we focus on the wireless charging energy in Section~\ref{charging_model} and motion energy consumption in this section. Based on this, we consider the use of rotary-wing AAV and SV equipped with DC motors, with their respective motion energy consumption models as follows:

\par For a rotary-wing AAV with a flight speed of $v$, its motion energy consumption~\cite{Zeng2019} can be given by

\begin{equation}
\begin{aligned}
    P_{AAV}(v)=&P_B\left(1+\frac{3v^2}{v_{tip}^2}\right)+\\
    &P_I\left(\sqrt{1+\frac{v^4}{4v_0^4}}-\frac{v^2}{2v_0^2}\right)^{1/2}+\frac12d_0\rho sAv^3,
\end{aligned}
\end{equation}

\noindent where $P_B$ and $P_I$ represent the blade power and induced power of the AAV in a hovering state, respectively. Moreover, $v_{tip}$ denotes the tip speed of the rotor blades, while $v_0$ represents the average induced rotor speed of the AAV in the hovering state. Additionally, $d_0$ and $\rho$ are the body drag coefficient and air density, respectively. Meanwhile, $s$ and $A$ represent the solidity and area of the rotor of the AAV, respectively.

\par For an SV with a travel speed of $v$ and using a permanent magnet direct current (PMDC) motor model, its motion energy consumption~\cite{Mei2004} can be given by

\begin{equation}
    P_{SV}(v)=k_1v^2+k_2v+k_3,
\end{equation}

\noindent where $k_1$, $k_2$, and $k_3$ are the respective control parameters.

\par Without loss of generality, we disregard the additional increase or decrease in energy consumption of the AAV and SV due to acceleration or deceleration during motion, as these account for only a small fraction of their total operating time.

% Subsection
% Problem Formulation and Analyses
%

\section{Problem Formulation and Analyses} \label{sec:problem_formulation_and_analysis}

\par In this section, we analyze the collaborative charging problem of HAGCCS. \textit{First}, we analyze several key factors involved in the charging phase. \textit{Second}, we formulate and analyze the collaborative charging problem.

\subsection{Problem Statement}

\par In this work, we focus on three optimization objectives, \textit{i.e.}, improving the charging efficiency of the AAV and SV, reducing the travel distance of the AAV and SV, and minimizing the mortality of the sensor nodes. These three optimization objectives involve inherent trade-offs. Specifically, if the AAV and SV are positioned closer to the sensor nodes, a larger number of nodes will fall within the charging range, thereby improving the charging efficiency. The location of the AAV and SV is directly related to their energy consumption, which means that if the positioning results in more frequent or longer travel of the AAV and SV, the energy consumption will increase accordingly. Moreover, improper positioning may lead to inadequate coverage of sensor nodes, thereby preventing some nodes from receiving sufficient charging support, which means that the node mortality increases.

\par As such, the corresponding decision variables are represented as $\boldsymbol{Z}_t=\{x^{SV}_t, y^{SV}_t, x^{AAV}_t, y^{AAV}_t, h^{AAV}_t\}$, whose variables correspond to the coordinates of the AAV and SV.

\par In HAGCCS, we aim to enhance the charging efficiency of the AAV and SV to supply more energy to the sensor nodes, thereby extending the lifetime of WRSN. According to Eq.~\eqref{eq:charging_limitation}, the AAV or SV can charge all sensor nodes within the effective charging radius $d_{max}$. Therefore, the charging efficiency of the AAV or SV, which is the first optimization objective $f_1$, can be expressed as follows:

\begin{equation}\label{eq:objective1}
    f_1=\sum_{i=1}^{N_{S}}P_i.
\end{equation}

\par By reducing the travel distance of the AAV and SV, the energy consumption caused by their travel distance is minimized. Therefore, more energy can be allocated for charging the sensor nodes, thereby effectively improving their energy utilization efficiency. Let $(x_{\text{init}}, y_{\text{init}}, z_{\text{init}})$ and $(x_{\text{target}}, y_{\text{target}}, z_{\text{target}})$ represent the initial and target positions of the AAV or SV, respectively, in a single movement, then the travel distance of the AAV or SV, which is the second optimization objective $f_2$, can then be expressed as follows:

\begin{equation}\label{eq:objective2}
    f_2=\sqrt{(x_{\text{target}}-x_{\text{init}})^2+(y_{\text{target}}-y_{\text{init}})^2+(z_{\text{target}}-z_{\text{init}})^2}.
\end{equation}

\par The mortality of sensor nodes is a key indicator for evaluating the performance and efficiency of WRSNs. Specifically, an increase in sensor node mortality leads to deterioration in WRSN stability and reliability, while also reducing the integrity of collected data. As such, we consider minimizing the mortality of sensor nodes in this network as the third optimization objective. Specifically, the third objective $f_3$, \textit{i.e.}, the mortality of sensor nodes,

\begin{equation}\label{eq:objective3}
    f_3=\frac{\sum_{i=1}^{N_{S}}b_i}{N_{\mathcal{S}}},
\end{equation}

\noindent where $b_i$ is a binary variable defined as follows:

\begin{equation}
b_i = 
\begin{cases}
0, & \text{if sensor node } i \text{ is alive} \\
1, & \text{if sensor node } i \text{ is dead}
\end{cases}.
\end{equation}

% R1-1

\noindent Note that $f_3$ is a shared network-level term applied to both the AAV and SV, whereas $f_1$ and $f_2$ are agent-specific terms that reflect the individual charging efficiency and travel distance of each agent. More importantly, $f_3$ is designed to encourage cooperative coverage between the AAV and SV and to incentivize both agents to coordinate proactively so as to minimize coverage gaps and maximize the overall lifetime.

% R2-4

\par To improve the charging efficiency, the AAV and SV need to move frequently between sensor nodes that need to be charged, which results in an increase in their travel distance. However, as the travel distances of the AAV and SV increase, their energy consumption also rises, which means that they cannot charge more sensors. As a result, the mortality of sensor nodes will increase. Therefore, three optimization objectives have a conflicting relationship. Thus, we formulate this problem by using multi-objective optimization theory.

\par According to the three optimization sub-objectives above, our optimization problem can be formulated as follows:

\begin{subequations}
    \label{eq:formulation1}
    \begin{align}
        (\mathrm{P1}): \quad {\underset{\boldsymbol{Z}_t}{\text{max}}} \sum^{\mathcal{T}}_{t=1}&(f_1, -f_2, -f_3), \\
        \text{s.t.} \quad \quad 
        & 0\leq x_t^{AAV}\leq X_{max},\quad \forall t\in \mathcal{T}\label{eq:formulation1_c1}\\
        & 0\leq y_t^{AAV}\leq Y_{max},\quad \forall t\in \mathcal{T}\label{eq:formulation1_c2}\\
        & 0\leq x_t^{SV}\leq X_{max},\quad \forall t\in \mathcal{T}\label{eq:formulation1_c3}\\
        & 0\leq y_t^{SV}\leq Y_{max},\quad \forall t\in \mathcal{T}\label{eq:formulation1_c4}
    \end{align}
\end{subequations}

\noindent where $X_{max}$ and $Y_{max}$ represent the maximum ranges of the WRSN area along the x-axis and y-axis, respectively. Moreover, the boundary constraints \eqref{eq:formulation1_c1}-\eqref{eq:formulation1_c2} and \eqref{eq:formulation1_c3}-\eqref{eq:formulation1_c4} ensure that both the AAV and SV operate within the WRSN boundaries, respectively.

\subsection{Problem Analyses}

% R2-1

\par Based on the HAGCCS and the optimization sub-objectives, problem (P1) exhibits the following characteristics. \textit{Firstly}, problem (P1) exhibits strong dynamic and stochastic characteristics. Specifically, the energy consumption magnitude of sensors varies randomly, thereby making the current overall network sensor energy consumption level unpredictable, which makes it challenging to dynamically capture critical state features, thus demonstrating strong dynamic properties. Moreover, given the limited energy budgets carried by the AAV and SV, the uncertainty in their travel distances makes their energy consumption stochastic, thereby reducing the energy available for charging the WRSN. \textit{Secondly}, this problem involves both long-term and short-term optimization objectives. In particular, the long-term objective is to maximize the WRSN lifetime, while the short-term objective is to minimize the energy consumption of the AAV and SV within each time slot. Therefore, during the optimization process, we should consider both the current and long-term interests. \textit{Finally}, since the AAV is an energy-sensitive system that requires real-time decision-making during flight operation, the solution used to solve this problem should satisfy real-time computational requirements.
% R1-2

\par Accordingly, the multi-objective optimization problem (P1) exhibits dynamic characteristics, long-term slot properties, and real-time decision-making requirements. Thus, conventional optimization methods or evolutionary computation algorithms are unsuitable for this problem. Specifically, conventional optimization methods typically rely on a known and fixed environment model~\cite{He2013}. Even if heuristic or evolutionary algorithms are used, they are often predefined or require a considerable amount of time to run, which prevents real-time adjustments in practical applications~\cite{Shu2016}. Moreover, these methods generally focus on immediate optimization and struggle to balance both current and long-term benefits. Though conventional methods may maximize short-term gains, they overlook the sustainability of long-term network performance and stability. Furthermore, due to the limited computing capability of their onboard devices and the constrained energy budgets of the AAV and SV, Pareto-based methods, which require high computational overhead to compute a set of nondominated solutions, are not suitable for solving this multi-objective problem. Therefore, a faster and more energy-efficient approach that better aligns with the goal of extending the WRSN lifetime is needed to solve problem (P1).
% R1-2

\par Accordingly, we adopt the advantageous DRL to address the considered problem. Specifically, DRL enables adaptive decision-making in dynamic environments and optimizes long-term network performance by learning from real-time feedback, thereby making it well-suited for problem (P1) in HAGCCS.

\section{Heterogeneous Trust Region Strategy Optimization-Based Decentralized Solution} \label{sec:solution}

\par In this section, we propose a decentralized solution to address the collaborative charging problem (P1) in the WRSN. \textit{Firstly}, we formulate the optimization problem as a Markov game (MG)~\cite{Gronauer2022} involving the AAV and SV agents. \textit{Secondly}, we introduce the IHATRPO algorithm that integrates a self-attention mechanism and Beta sampling to enhance multi-agent coordination. \textit{Finally}, we analyze the computational and space complexity of the proposed algorithm.

%
% Markov game in IHATRPO
%
\subsection{Markov Game Formulation}

\par We first model Problem (P1) as a MG. Specifically, MG can be formally represented by the tuple $\langle\allowbreak \mathcal{N},\allowbreak \{\mathcal{S}_i\}_{i \in \mathcal{N}},\allowbreak \{\mathcal{A}_i\}_{i \in \mathcal{N}},\allowbreak \mathcal{P},\allowbreak \{\boldsymbol{R}_i\}_{i \in \mathcal{N}},\allowbreak \gamma \rangle$. The key elements of MG are given as follows:
% R3-6

\subsubsection{Agent Set} 

\par The HAGCCS employs two agents that are assigned to control the AAV and SV, respectively, \textit{i.e.},
\begin{equation}\label{eq:agents}
    \mathcal{N}=\{A^{AAV},A^{SV}\}.
\end{equation}

\par At each time slot $t$, both agents independently observe the environmental state and execute actions, so as to maximize their respective expected total rewards.

\subsubsection{State Space} 
Both agents share the same global state space, which can ensure complete environmental observability for decision-making. Specifically, the state space consists of the energy levels and positions of sensor nodes, which can be collected by BS via existing WRSN communication protocols, and positions of the AAV and SV, which can be acquired via the global positioning system (GPS). Moreover, the acquisition of such state information through the inherent communication protocols in WRSNs incurs no significant additional communication overhead. Therefore, the state space is defined as follows:
% R1-3

\begin{equation}\label{eq:state_space2}
    \mathcal{S}=\{s_t|s_t=(\mathcal{S}_t,AAV_t,SV_t), \forall t \in \mathcal{T}\},
\end{equation}
\noindent where $\mathcal{S}_{t} = \{x_{t}^{1},\allowbreak x_{t}^{2},\allowbreak ...,\allowbreak x_{t}^{N_{\mathcal{S}}},\allowbreak y_{t}^{1},\allowbreak y_{t}^{2},\allowbreak ...,\allowbreak y_{t}^{N_{\mathcal{S}}},\allowbreak q_{t}^{1},\allowbreak q_{t}^{2},\allowbreak ...,\allowbreak q_{t}^{N_{\mathcal{S}}}\}$ represents the set of coordinates, and current energy levels of each sensor node at the beginning of time slot $t$. Meanwhile, $AAV_t=\{x_t^{AAV},y_t^{AAV}, h^{AAV}\}$ and $SV_{t}=\{x_{t}^{SV},y_{t}^{SV}\}$ denote the coordinates of the AAV and SV, respectively, at the start of time slot $t$.

\subsubsection{Action Space} 

Each agent operates within its own action space, representing distinct decision variables for controlling vehicle motion parameters. Both the AAV and SV agents follow the same mathematical formulation while maintaining independent control over their respective vehicles. Based on environmental observations, each agent governs two critical motion parameters, which are the heading angle $\theta$ and travel distance $d$. Note that these two parameters can correspond to the decision variable $\boldsymbol{Z}_t$ of problem (P1). Consequently, the action space for each agent is defined as:
\begin{equation}\label{eq:action_space2}
    \mathcal{A}_i=\{a^{i}_{t}|a^{i}_{t}=(\theta^{i}_{t}, d^{i}_{t}), \forall t \in \mathcal{T}, i \in \mathcal{N}\}.
\end{equation}

\subsubsection{Reward Function} 

The reward mechanism is designed to motivate both agents to optimize their respective contributions to the HAGCCS performance. Each agent receives individual rewards based on its performance, with both agents sharing the same mathematical reward structure to ensure consistency and fairness in the learning process. Specifically, we combine three optimization sub-objectives, \textit{i.e.}, charging efficiency, energy consumption represented by travel distance, and network sustainability measured by node mortality, into a scalarized reward function through a weighted sum, which enables faster solution finding in WRSN charging scenarios with limited computing resources and energy budgets. The reward function is defined as follows:

% R1-2
% According to the optimization objectives in problem (P1), the reward function incorporates three key performance indicators: charging efficiency, energy consumption (represented by travel distance), and network sustainability (measured by node mortality). The reward function is defined as follows:
\begin{equation}\label{eq:reward_function}
    \mathcal{R}_i = \{r^{i}_{t}|r^{i}_{t}=\lambda_1f^i_{1,t}-\lambda_2f^i_{2,t}-\lambda_3f_{3,t}, \forall t \in \mathcal{T}, i \in \mathcal{N}\},
\end{equation}

\noindent where $f^i_{1,t}$, $f^i_{2,t}$, and $f_{3,t}$ correspond to Eq.~\eqref{eq:objective1}, Eq.~\eqref{eq:objective2}, and Eq.~\eqref{eq:objective3} during time slot $t$. The weighting coefficients $\lambda_1$, $\lambda_2$, and $\lambda_3$ serve as balancing factors that ensure appropriate emphasis on the relative importance of each reward component in the overall system performance. Furthermore, we normalize the three sub-objectives during the weighted sum process using $\lambda_1$, $\lambda_2$, and $\lambda_3$, respectively, whose values will be further analyzed in Section~\ref{sec:weightanalysis}.
% R1-2

% Algorithm
% IHATRPO
%
\subsection{IHATRPO Algorithm}

\begin{algorithm}[t]
    \normalem
    \small
    \caption{IHATRPO}
    \label{algo:IHATRPO}

    \KwIn{Number of heterogeneous agents $n$, Max training episodes $max\_episodes$, max time slots $max\_time\_slots$}
    \KwOut{Optimized policy network parameters $\{\theta_i\}_{i=1}^n$}

    \tcc{Initialization:}

    \For{agent $i \in [1,n]$}
    {
        Initialize Actor network parameters $\theta_i$ and Critic network parameters $\omega_i$
    }

    \For{$episode$ = $1$ to $max\_episodes$}
    {
        Reset sensor nodes, initialize power levels for AAV/SV, initialize trajectory buffer $\mathcal{D}$
        
        \For{$t$ = $1$ to $max\_time\_slots$}
        {
            \For{agent $i \in [1,n]$}
            {
                Agent $i$ constructs Beta distribution from state $s_t$, samples action $a_i^t$

                Execute action $a_i^t$, receive reward $r_i^t$
            }
             Update environment state $s_t \rightarrow s_{t+1}$

            \For{agent $i \in [1,n]$}
             {
                Store transition $(s_t, a_i^t, s_{t+1}, r_i^t)$ in trajectory buffer $\mathcal{D}$
             }

            \If{ $E_{AAV} \leq 0$ and $E_{SV} \leq 0$}
            {
                break
            }
        }
        \tcc{Policy Update:}
        \For{agent $i \in [1,n]$}
        {
            Compute generalized advantage estimation (GAE) advantages $\hat{A}^{\pi_{\theta_i}}$ from $\mathcal{D}$ and normalize

            Update $\omega_i$

            Update $\theta_i$ via TRPO using Eq.~\eqref{eq:parameterOptimizationOriginal}
        }
    }
    \textbf{Return} $\theta = \{\theta_1, \dots, \theta_n\}$.
\end{algorithm}

\par In this section, we handle the MG through the IHATRPO algorithm, where the AAV and SV are each treated as an agent. In the following, we first introduce the conventional HATRPO. Subsequently, we present two improvement measures, namely a self-attention mechanism and Beta sampling, to enhance the ability of HATRPO to handle the MG. 

\subsubsection{Preliminaries of HATRPO}\label{sec:HATRPOintroduction}

\par HATRPO integrates the multi-agent framework with trust region policy optimization to enhance multi-agent DRL (MADRL), thus achieving monotonic improvement.

\par In an $N$-agent MG, the joint policy $\boldsymbol{\pi}=(\pi_{1},\ldots,\pi_{N})$ represents collective decision-making of agents. Specifically, at time slot $t$, given state $s^t$, each agent takes an action $a^t_i$ according to its policy. Subsequently, the environment computes the reward $\boldsymbol{r}^t=(r^t_1,\ldots,t^t_N)$ based on the joint action $\boldsymbol{a}^t=(a^t_1,\ldots,a^t_N)$ and updates the state to $s^{t+1}$. The optimization goal is to maximize expected cumulative reward by updating policy parameters from $\theta_i$ to $\theta^{\prime}_{i}$, where the objective function difference from the policy update is given by
\begin{equation}\label{eq:objective_function_difference}
    J(\theta_i^{\prime})-J(\theta_i) = \mathbb{E}_{\tau \sim \boldsymbol{\pi}} \left[ \sum_{t=0}^{\infty} \gamma^t A^{\pi_{\theta_i}}(s^t_i, a^t_i) \right],
\end{equation}

\noindent where $\tau$ is the trajectory, $\gamma \in (0,1)$ is the discount factor, and $A^{\pi_{\theta_i}}$ is the advantage function under policy $\pi_{\theta_i}$. However, since the updated policy $\pi_{\theta_{i}^{\prime}}$ cannot be computed directly, we approximate the objective function using the state distribution of the pre-update policy $\pi_{\theta_i}$ and apply importance sampling to correct the action distribution. The objective is then given by
\begin{equation}L(\theta_{i}^{\prime}|\theta_{i})=\mathbb{E}_{s_i \sim \nu^{\boldsymbol{\pi}}}\mathbb{E}_{a_i \sim \pi_{\theta_i}(\cdot|s_i)}\left[ \frac{\pi_{\theta_{i}^{\prime}}(a_i|s_i)}{\pi_{\theta_i}(a_i|s_i)}A^{\pi_{\theta_i}}(s_i, a_i) \right].
\end{equation}

\par To maintain proximity between the updated and original policies, we adopt the Kullback-Leibler (KL) divergence within the trust region policy optimization framework~\cite{Schulman2015}. Specifically, the divergence between the pre-update policy $\pi_{\theta_i}$ and post-update policy $\pi_{\theta_i^{\prime}}$ is denoted by $D_{KL}(\pi_{\theta}||\pi_{\theta_{i}^{\prime}})$. By setting $\delta$ as the update step size threshold, we formulate the optimization problem as:

\begin{equation}\label{eq:HATRPO_optimization_formula}
\begin{aligned}
    &\text{max}_{\theta_i^{\prime}} \quad  L(\theta_{i}^{\prime}|\theta_{i}) \\
    &\text{s.t.} \quad \mathbb{E}_{s_i \sim \nu^{\boldsymbol{\pi}}}\left[D_{KL}(\pi_{\theta_i}||\pi_{\theta_i^\prime})\right] \leq \delta.
\end{aligned}
\end{equation}

\par To simplify the computation, we apply linear and quadratic approximations to the objective function and KL constraint, respectively, thereby yielding the closed-form update as follows:
\begin{equation}\label{eq:parameterOptimizationOriginal}
    \theta^{k+1}_i=\theta_i^k+\alpha^j\sqrt{\frac{2\delta}{(g_i)^T(H_i)^{-1}g_i}}{H_i}^{-1}g_i,
\end{equation}

\noindent where $\theta^k_i$ represents the policy parameters after the $k$-th iteration of the $i$-th agent, $\alpha^j\in (0, 1)$ is the backtracking line search coefficient, which ensures that $\theta_i^{\prime}$ is superior to $\theta^k_i$ and satisfies the KL divergence constraint. Moreover, $g_i=\nabla_{\theta_i^\prime}\mathbb{E}_{s_i \sim \nu^{\boldsymbol{\pi}}}\mathbb{E}_{a_i \sim \pi_{\theta_i^k}(\cdot|s_i)}[\pi_{\theta_i^\prime}(a_i|s_i)/\pi_{\theta_i^k}(a_i|s_i)A^{\pi_{\theta_i^k}}(s_i, a_i)]$ is the gradient of the optimization objective, and $H_i=\mathcal{H}\left[\mathbb{E}_{s_i\sim\nu^{\boldsymbol{\pi}}}\left[D_{KL}\left(\pi_{\theta_i}||\pi_{\theta_i^\prime}\right)\right]\right]$ represents the Hessian matrix derived from the KL divergence.

\subsubsection{Self-Attention Mechanism}

\par In HAGCCS, heterogeneous charging agents must simultaneously process multi-dimensional state information, including their own states, distributed sensor node conditions, and inter-agent coordination requirements within a non-stationary environment. Conventional MADRL approaches treat all state information equally through conventional feature extraction, thereby failing to capture varying feature importance and dynamic relationships between the AAV and SV and sensor nodes, which leads to suboptimal decision-making.
% R2-1

\par The self-attention mechanism addresses these limitations by dynamically assigning importance weights to input elements based on contextual relevance. Different from traditional approaches, the self-attention mechanism captures complex dependencies through parallel processing while adaptively focusing on critical information for decision-making. In our IHATRPO, we integrate the self-attention mechanism into the actor networks of both AAV and SV agents for heterogeneous multi-agent coordination. Specifically, the self-attention mechanism~\cite{Vaswani2017} computes context-aware representations by measuring similarity between input elements using Query ($Q$), Key ($K$), and Value ($V$) vectors, which can be given by
% R3-2
\begin{equation}
    A(Q, K, V)=\sigma\left(\frac{QK^T}{\sqrt{d_k}}\right)V,
\end{equation}

\noindent where $d_k$ represents the dimension of $K$. By using the self-attention mechanism integration, heterogeneous agents dynamically prioritize relevant information based on context and achieve a deep understanding of state interdependencies for informed decision-making. Note that since the policy is trained offline and deployed in a fixed network topology, this complexity does not constitute a significant bottleneck during deployment.
% R2-2

\subsubsection{Beta Sampling}

\begin{figure}[htbp]
    \centering
    \includegraphics[width=1\linewidth]{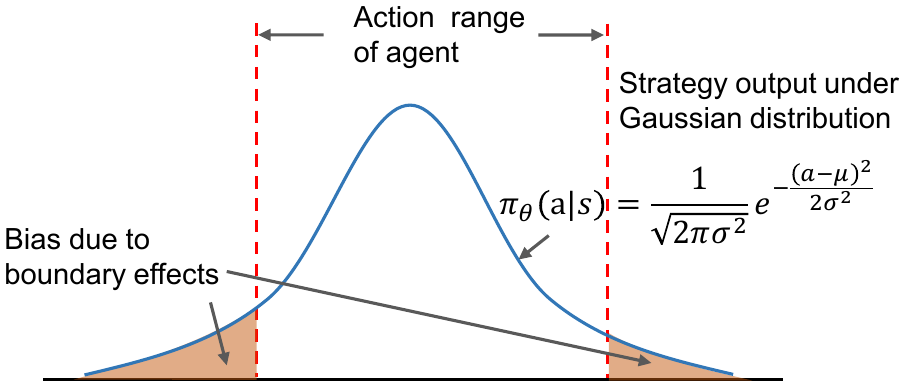}
    \caption{Boundary effects on Gaussian distribution bias. The shaded areas represent probability mass falling outside the valid action range, which must be truncated during sampling.}
    \label{fig:GaussianBias}
\end{figure}

\par The HAGCCS requires continuous action control for the AAV and SV travel within bounded action spaces constrained by the finite distribution range of the WRSN. However, conventional continuous control methods utilize the Gaussian distribution for action sampling, whose unbounded nature conflicts with the bounded action spaces in our work, thereby resulting in boundary effects and distributional bias that compromise gradient computation accuracy, as demonstrated in Fig.~\ref{fig:GaussianBias}.
% R4-2

\par In this case, the Beta distribution addresses these limitations through its inherent bounded property on $\left[0, 1\right]$, thereby ensuring all sampled actions remain within valid ranges without truncation. Unlike the Gaussian distribution that requires clipping or rescaling, thereby resulting in computation bias, Beta distributions naturally maintain unbiased gradient computation while respecting action space constraints~\cite{Chou2017}. The probability density function of the Beta distribution is given by
% R4-2
\begin{equation}
    f(x; \alpha_b, \beta_b)=\frac{\Gamma(\alpha_b+\beta_b)}{\Gamma(\alpha_b)\Gamma(\beta_b)}x^{\alpha_b-1}(1-x)^{\beta_b-1},
\end{equation}
\noindent where $\Gamma(\cdot)$ is the Gamma function, and $\alpha_b$ and $\beta_b$ serve as shape parameters that collectively determine the distribution shape. We adopt $\pi_{\theta}(a|s)=f(c\cdot a;\alpha_b, \beta_b)$ to characterize the stochastic policy, which is referred to as the Beta sampling strategy. The parameters $\alpha_b=\alpha_{b,\theta}(s)$ and $\beta_b=\beta_{b,\theta}(s)$ are modeled by a neural network parameterized by $\theta$. The parameter $c$ is determined based on the value ranges of travel direction and distance for the AAV or SV in the action space, thereby ensuring that action outputs satisfy their respective action space constraints.

\begin{figure*}[ht]
    \centering
    \includegraphics[width=1\linewidth]{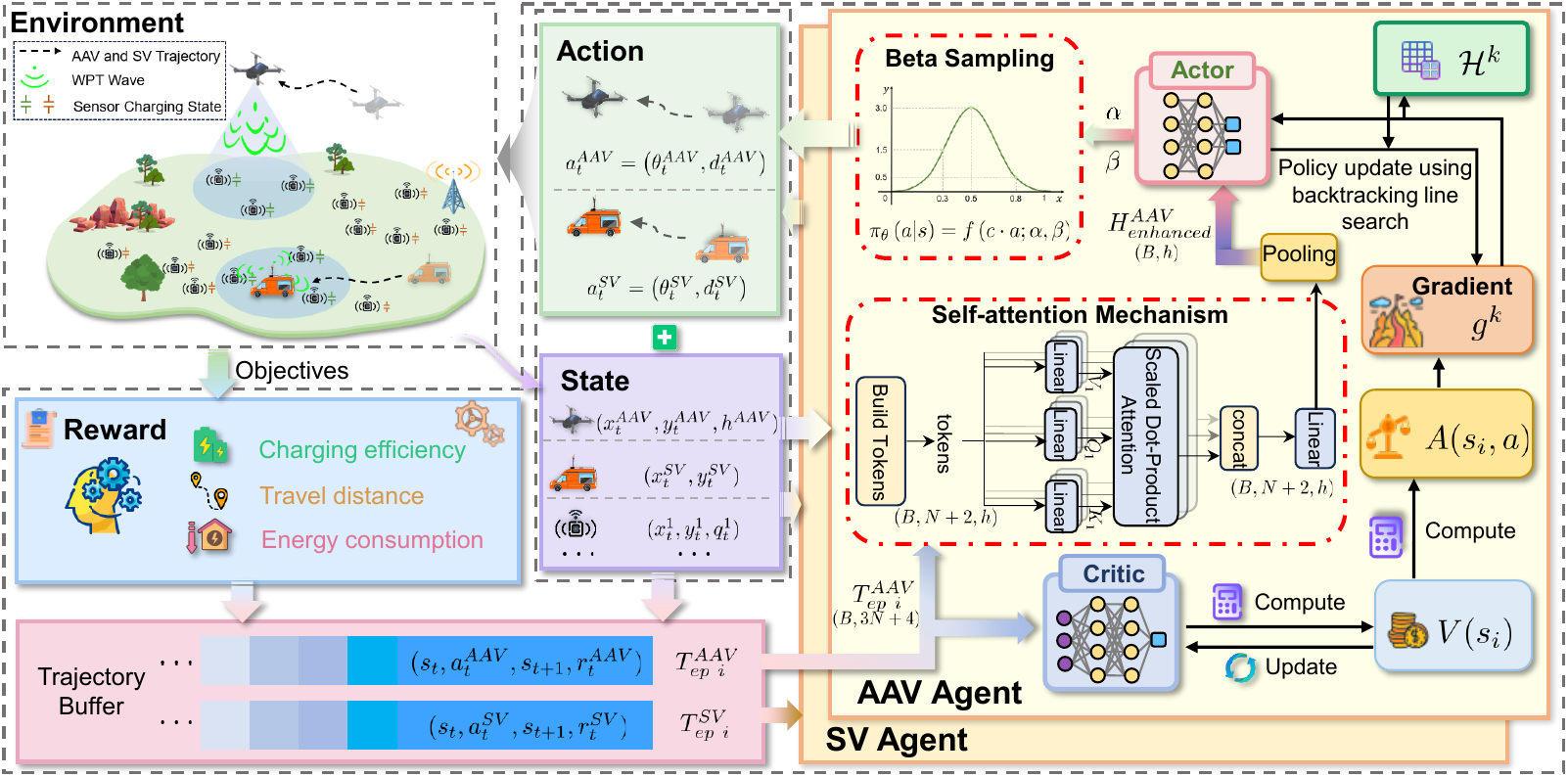}
    \caption{Framework of IHATRPO for heterogeneous air-ground collaborative charging in the WRSN. The algorithm integrates Beta distribution-based action sampling, self-attention mechanism-enhanced state processing in actor networks for the AAV and SV to optimize multi-objective charging strategies.}
    \label{fig:IHATRPO_algo}
\end{figure*}

% \FloatBarrier
\par Through Beta sampling implementation, the agents maintain unbiased gradient computation within bounded action spaces, eliminate boundary effects that degrade policy performance, and ensure natural action space compliance without additional constraints or post-processing steps.

\subsection{Main Steps and Complexity Analysis of IHATRPO}

\par Based on the methods introduced above, we propose the IHATRPO framework to address the multi-objective problem (P1). The overall structure of IHATRPO is depicted in Fig.~\ref{fig:IHATRPO_algo}, and the detailed procedure is outlined in Algorithm~\ref{algo:IHATRPO}.
\par As shown in Fig.~\ref{fig:IHATRPO_algo}, IHATRPO treats the AAV and SV as two independent agents that are trained and executed separately. Specifically, in each time slot, the AAV and SV agents each receive the global state of the environment and extract key features through the self-attention mechanism. Based on these features, each agent samples a bounded continuous action from the Beta sampling strategy to determine its movement direction and travel distance for the next step. Following this sequential decision process, rewards generated for the AAV and SV agents from the environment and state transitions are stored in the trajectory buffer. At the end of each episode, the collected trajectories are used to update the critic networks and optimize the actor networks through trust-region constrained policy updates for the AAV and SV agents. By repeating this process, IHATRPO gradually optimizes the AAV and SV as heterogeneous agents that collaboratively provide charging services for the WRSN.
\par More notably, the IHATRPO framework follows the centralized training and centralized execution paradigm because both the AAV and SV agents require the global state information, including the energy levels and positions of all sensor nodes as well as positions of the AAV and SV, so as to coordinate effectively in providing wireless charging services for the WRSN.

% R1-3

\par We analyze the computational and space complexity of the IHATRPO algorithm during both the training and execution phases. The computational complexity of IHATRPO during the training phase is $\mathcal{O}(N_A(|\boldsymbol{\theta}|+|\boldsymbol{\omega}|+N_E(T(V+N^2h)+|\boldsymbol{\omega}|+N_T(N_K+N_MN^2h)+|\boldsymbol{\theta}|(N_K+N_MN^2h)))$, which can be summarized as follows:

\begin{itemize}

    \item \textit{Network Initialization:} This phase involves the initialization of network parameters of the AAV and SV. Specifically, the computational complexity is expressed as $\mathcal{O}(N_A(|\boldsymbol{\theta}| + |\boldsymbol{\omega}|))$, where $N_A$ is the number of agents, the $| \cdot | $ operation represents the number of parameters in the networks.
    
    \item \textit{Action Selection:} This phase entails selecting actions according to feature extraction of the self-attention mechanism and the output scores of the actor network, and corresponding complexity is $\mathcal{O}(N_AN_ETN^2h)$, where $N$ is the number of sensor nodes and $h$ is the embedding dimension of the self-attention mechanism. Moreover, $N_E$ denotes the number of training episodes, and $T$ is the number of steps per episode.
% R4-4
    \item \textit{Reward Calculation and State Transitions:} The computational complexity of reward calculation and state transitions is $\mathcal{O}(N_AN_ETV)$, where $V$ represents the complexity of interacting with the environment.
    
    \item \textit{Network Update:} The updating phase consists of two main parts that are the updates of the critic networks, as well as the updates of the actor networks. \textit{First}, the advantage function is calculated, and the critic network parameters are updated subsequently. This part has the complexity of $\mathcal{O}(N_AN_E(|\boldsymbol{\omega}|+N_T))$, where $N_T$ is the length of the sampled training data. \textit{Second}, the actor network is updated by calculating the target value of the surrogate function, calculating the conjugate gradient, and linearly searching for parameters that meet the conditions. Therefore, the corresponding complexity is $\mathcal{O}(N_AN_E(N_T(N_K+N_MN^2h)+|\boldsymbol{\theta}|(N_K+N_MN^2h)))$, where $N_K$ is the number of iterations of the conjugate gradient and $N_M$ is the number of iterations of the linear search. Thus, the complexity of this phase is calculated as $\mathcal{O}(N_AN_E(|\boldsymbol{\omega}|+N_T(N_K+N_MN^2h)+|\boldsymbol{\theta}|(N_K+N_MN^2h)))$.
% R4-4
\end{itemize}
\par Note that, as for the computational complexity, the network update phase dominates the computational complexity during training, particularly the processes of conjugate gradient computation and linear search combined with the attention mechanism for optimizing the actor network parameters.
% R4-4

\par Besides, the space complexity of IHATRPO during the training phase is $\mathcal{O}(N_A(|\boldsymbol{\theta}| + |\boldsymbol{\omega}|) + |\mathcal{D}|(|\boldsymbol{s}|+\boldsymbol{a})$, where $|\mathcal{D}|$ denotes the size of the trajectory buffer. As such, the space complexity is mainly for storing neural network parameters and sampled trajectories.
% R4-4
\par During the evaluation phase, the computational complexity of IHATRPO is $\mathcal{O}(N_AN_EN^2h)$, which can be attributed to action selection and transition according to the current state using the feature and actor network. Moreover, the space complexity during the execution phase is $N_A|\boldsymbol{\theta}|$ since the feature and actor network parameters need to be stored in memory for action selection.

%
%Simulations and Analyses
%
\section{Simulations and Analyses} \label{sec:simulation_results_and_analysis}

\par In this section, we first introduce the simulation settings and baselines. Subsequently, we present the optimization results, the comparison analyses with state-of-the-art baselines, and the analysis of agent spatial movement patterns.

\subsection{Simulation Setups}

\subsubsection{Scenario and Algorithm Setups}

\par In the simulations, we consider the scenario that the AAV and SV provide wireless charging to a sensor network. The primary parameters are shown in Table~\ref{tab:setups}. Additionally, following the methodology in~\cite{Fu2016}, we set the charging efficiency parameters $\alpha$ and $\beta$ in Eq.~\eqref{eq:charging_limitation} to 36 and 30, respectively. The energy consumption rate of sensor nodes per round is randomly generated within the range of 0.025 J to 0.04 J.

% R3-2
\par In the proposed IHATRPO, the algorithm parameters are shown in Table~\ref{tab:setups}. Both the policy network and value network are configured with two hidden layers, each containing 256 neurons. Meanwhile, in the self-attention mechanism, the number of heads is set to 4, and the embedding dimension is set to 256. Additionally, we set the number of training iterations to $6.5\times10^5$ and employ the Adam optimizer for neural network updates. Note that these algorithm parameters are determined by careful tuning to ensure performance and convergence. We consider the heterogeneity between the AAV and SV by assigning different reward weight coefficients, which are determined experimentally and shown in Table~\ref{tab:setups}.

\begin{table}[h]
\centering% This centers the table
\caption{Simulation settings} % This adds a title to the table
\label{tab:setups}
\begin{tabularx}{3.5in}{p{5.5cm}p{2.5cm}}
\toprule
\textbf{Parameters}                                    & \textbf{Values}                     \\\midrule
Network area                                            & 100 $\times$ 100 $m^2$                \\
Number of sensor nodes                                            & 100                    \\
Transmit power of AAV and SV                          & 3 W~\cite{Hou2024}             \\
Reception threshold of the sensor node                               & 5 mW           \\
The max energy of the sensor node                                    & 2 J                \\
The charging radius of AAV and SV                       & 6 m                           \\
Learning rate of neural network                         & $5\times10^{-5}$                      \\
KL threshold                                            & $5\times10^{-5}$                 \\
Linear search step                                      & 0.5                            \\
GAE scaling factor $\lambda$                            & 0.98                           \\
Entropy coefficient                                     & 0.01                            \\
Discount factor                                        & 0.96                             \\
Time step of each episode                               & 200                             \\
$\lambda_1$, $\lambda_2$, and $\lambda_3$ for AAV       &1, 0.001, and 1         \\
$\lambda_1$, $\lambda_2$, and $\lambda_3$ for SV      &1, 0.03, and 0.1        \\
\bottomrule
\end{tabularx}
\end{table}

\begin{figure*}[h]
\begin{minipage}[t]{1\linewidth}
  \centering
	\subfloat[]{\includegraphics[width=.24\linewidth]{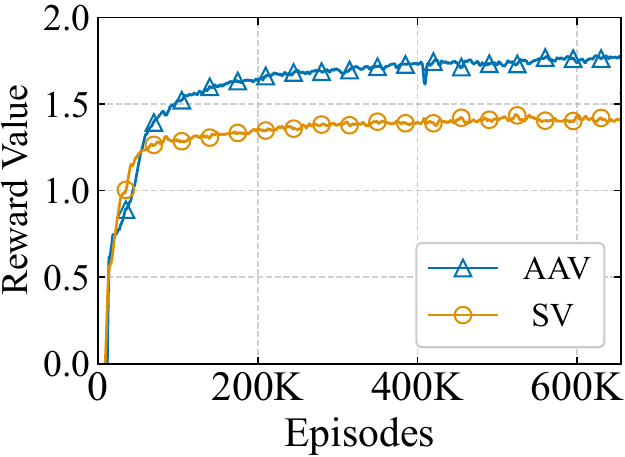}}\hfill
        \subfloat[]{\includegraphics[width=.24\linewidth]{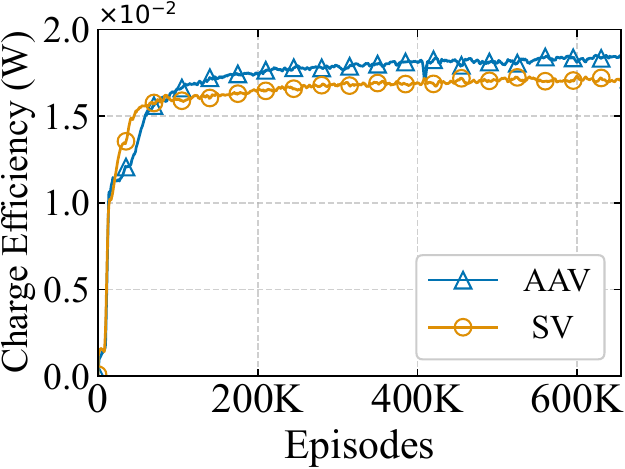}}\hfill
	\subfloat[]{\includegraphics[width=.24\linewidth]{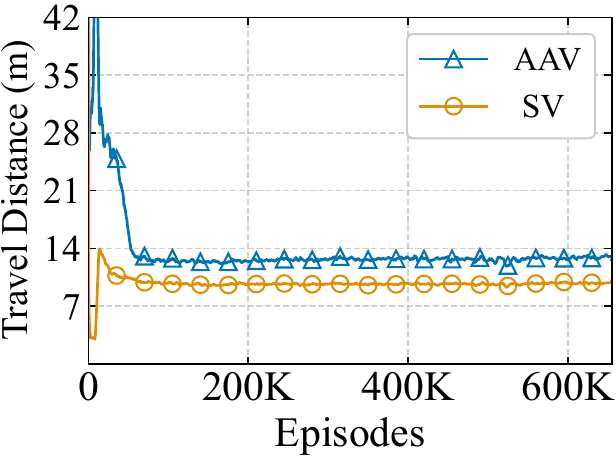}}\hfill
	\subfloat[]{\includegraphics[width=.24\linewidth]{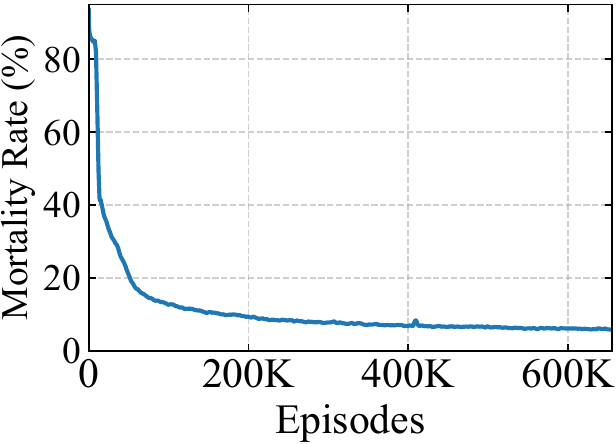}}\hfill
    \caption{Visualization results obtained by IHATRPO. (a) The total reward of the AAV and SV, respectively. (b) The charging efficiency of the AAV and SV. (c) The travel distance of the AAV and SV. (d) The mortality of sensor nodes ($100\times f_3 \%$).}
    \label{fig:convergences}
\end{minipage}
\end{figure*}

\subsubsection{Baselines}

\par To demonstrate the superiority of the proposed IHATRPO, we introduce the following comparative baselines. Note that these baselines adopt the same parameters as mentioned above and integrate the schedule policy of the AAV and SV.

\begin{itemize}
    \item \textit{PPO}: PPO is a policy gradient method that improves training stability through clipped surrogate objectives~\cite{Schulman2017}. As a single-agent baseline, PPO treats the multi-agent environment as a stationary single-agent MDP by training each agent independently, therefore ignoring the non-stationary nature caused by other learning agents.

    \item \textit{DDPG}: DDPG is an actor-critic method designed for continuous control tasks that combines policy gradient methods with Q-learning~\cite{Lillicrap2016}. When applied to multi-agent settings, each agent is trained independently using DDPG and treats other agents as part of the environment dynamics without explicit coordination mechanisms.
    
    \item \textit{MADDPG}: MADDPG is a DDPG-based classical MADRL approach based on the centralized training and decentralized execution architecture~\cite{Lowe2017}. This baseline allows agents to access global information during training while maintaining individual policies and shows effectiveness in multi-agent continuous control tasks.

    \item \textit{Heterogeneous-Agent PPO (HAPPO)}: HAPPO adapts PPO for heterogeneous multi-agent settings where agents have different observation and action spaces~\cite{Kuba2022}. This method serves as a baseline given its capability to handle heterogeneous AAV-SV coordination.

    \item \textit{HATRPO}: HATRPO extends TRPO to a heterogeneous multi-agent environment by maintaining individual trust regions for each agent~\cite{Kuba2022}. Furthermore, the details of this approach are elaborated in Section~\ref{sec:HATRPOintroduction}. The implementation ensures stable policy updates through KL divergence constraints across diverse agents. 
\end{itemize}

\par As such, the comparisons with PPO and DDPG demonstrate the necessity of multi-agent coordination mechanisms, the comparison with MADDPG shows the effectiveness of handling different types of agents, the comparison with HAPPO illustrates the superiority of the HATRPO-based framework in handling heterogeneous multi-agent scenarios, and the comparison with HATRPO can assess the effectiveness of two improvement measures of IHATRPO. In the following analyses, we first present the performance of multiple optimization sub-objectives under the IHATRPO, and then conduct a comparative analysis of convergence performance and total reward feedback between these baselines and IHATRPO, and the following analysis of agent trajectories.

\subsection{Performance Evaluation}

\subsubsection{Optimization Results}
\label{sec:weightanalysis}

% 这里修改了d图的表述$100\times f_3 \%$ 而不是$f_3$.    R3-4
\par Fig.~\ref{fig:convergences}(a) shows the respective cumulative reward of the AAV and SV, Fig.~\ref{fig:convergences}(b), Fig.~\ref{fig:convergences}(c), and Fig.~\ref{fig:convergences}(d) illustrate optimization of objectives in terms of the charging efficiency ($f_1$), travel distance ($f_2$) of the AAV and SV, and the mortality rate ($100\times f_3 \%$) of sensor nodes. As can be seen, the AAV and SV agents exhibit similar convergence trends and converge after approximately 200k iterations in Fig.~\ref{fig:convergences}(a), which demonstrates that IHATRPO, which employs the heterogeneous optimization framework, achieves good optimization performance for heterogeneous agents. Moreover, each objective achieves good optimization results with increasing training episodes in Figs.~\ref{fig:convergences}(a), (b), and (c), which demonstrates that the proposed reward function in Eq.~\eqref{eq:reward_function} can better balance the relationship between the AAV and SV. Moreover, it is noteworthy that a significant reduction in sensor node mortality from an initial rate exceeding 90\% to below 10\% in Fig.~\ref{fig:convergences}(d), which indicates that through the scheduling of the AAV and SV, the sensor node mortality can be reduced and HAGCCS achieves better energy efficiency.

\par
\begin{figure*}[!t]
\begin{minipage}[t]{1\linewidth}
  \centering
	\subfloat[]{\includegraphics[width=.19\linewidth]{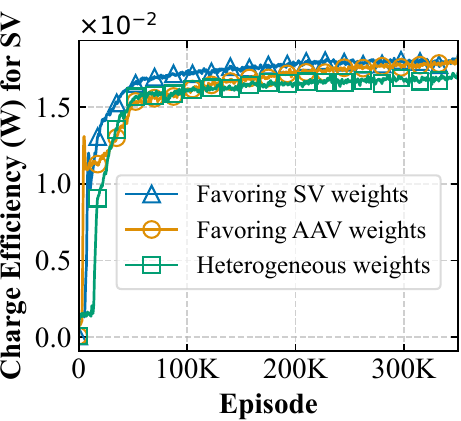}}\hfill
        \subfloat[]{\includegraphics[width=.19\linewidth]{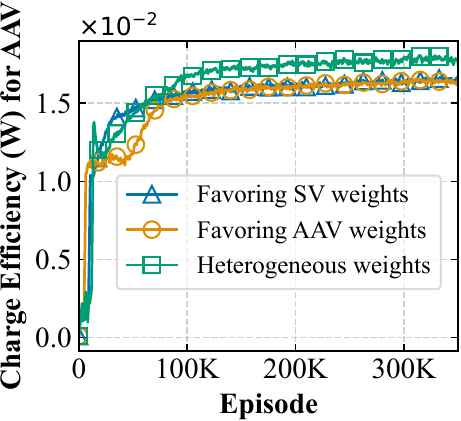}}\hfill
	\subfloat[]{\includegraphics[width=.19\linewidth]{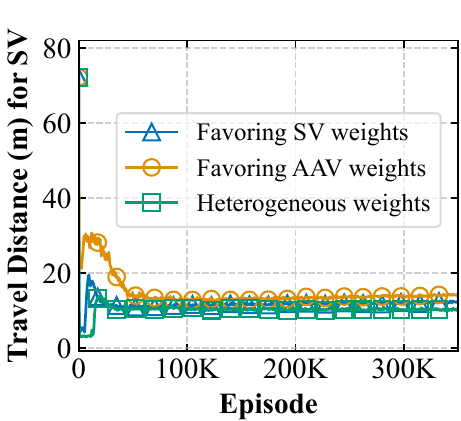}}\hfill
	\subfloat[]{\includegraphics[width=.19\linewidth]{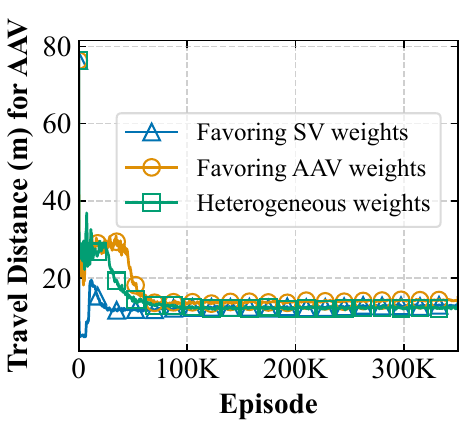}}\hfill
    \subfloat[]{\includegraphics[width=.19\linewidth]{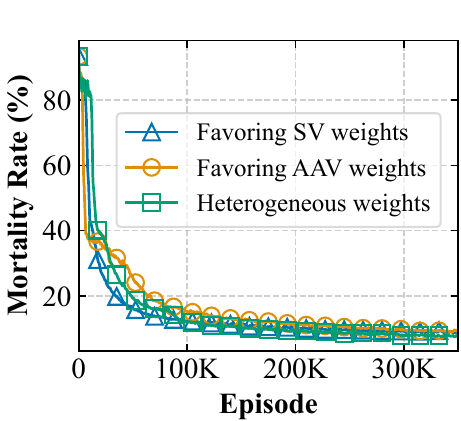}}\hfill
    \caption{Results of IHATRPO under different weights of the reward function. (a) The charging efficiency for SV. (b) The charging efficiency for AAV. (c) The travel distance for SV. (d) The travel distance for AAV. (e) The mortality of sensor nodes ($100\times f_3 \%$).}
    \label{fig:weight}
\end{minipage}
\end{figure*}

 Furthermore, to validate the heterogeneous weight assignment, we compare three configurations in Fig.~\ref{fig:weight}. As shown in Fig.~\ref{fig:weight}, the proposed heterogeneous assignment achieves more stable convergence and optimization results on the travel distance and the mortality rate of sensor nodes, thereby confirming the necessity of agent-specific weight configuration.

\begin{figure}[htbp]
    \centering
    \includegraphics[width=0.9\linewidth]{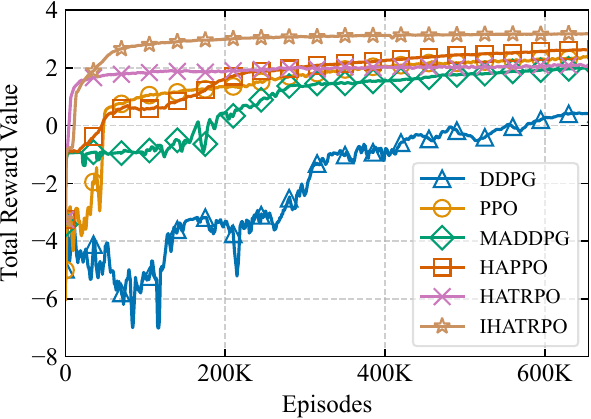}
    \caption{Convergence performance comparison of PPO, DDPG, MADDPG, HAPPO, HATRPO, IHATRPO.}
    \label{fig:mutli-algorithm}
\end{figure}

\subsubsection{Comparison Results}

\par Fig.~\ref{fig:mutli-algorithm} illustrates the cumulative rewards for each episode of IHATRPO in comparison to other benchmark algorithms. As can be seen, IHATRPO achieves faster convergence speed and the highest reward. This can be explained by several factors. \textit{First}, the self-attention mechanism enables IHATRPO to dynamically prioritize relevant information and capture complex dependencies among multi-dimensional states. \textit{Second}, the Beta sampling provides naturally bounded action sampling complying with the HAGCCS characteristics for IHATRPO. While HATRPO demonstrates the fastest convergence performance in the initial phase, this algorithm achieves lower cumulative rewards after convergence due to its limited capability in processing high-dimensional information and coordinating between heterogeneous agents. Among the single-agent baselines, PPO shows better convergence than DDPG, but both struggle with the multi-agent coordination challenges for convergence. MADDPG fails to account for the heterogeneity between the AAV and SV, thereby resulting in slower convergence and lower reward. HAPPO and HATRPO exhibit slower convergence due to action boundary violations caused by Gaussian sampling.

\par The performance improvement over the original HATRPO particularly validates that the integration of the self-attention mechanism and Beta sampling strategy effectively enhances policy optimization capability, thereby achieving superior learning performance in the heterogeneous multi-agent collaborative charging scenario.

\subsubsection{Sensitivity Analysis}

\begin{figure*}[!t]
\begin{minipage}[t]{1\linewidth}
  \centering
	\subfloat[]{\includegraphics[width=.19\linewidth]{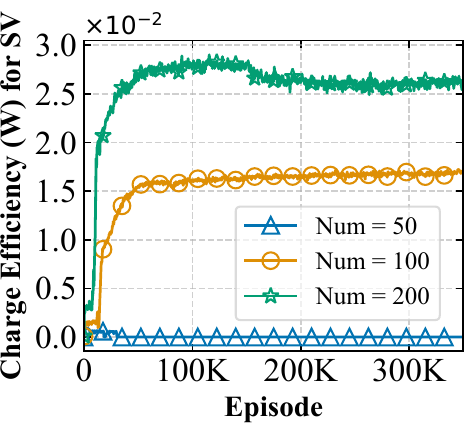}}\hfill
        \subfloat[]{\includegraphics[width=.19\linewidth]{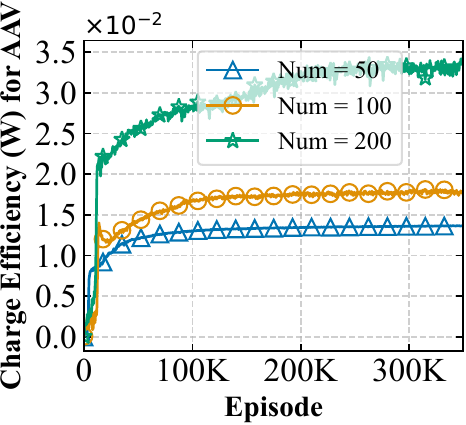}}\hfill
	\subfloat[]{\includegraphics[width=.19\linewidth]{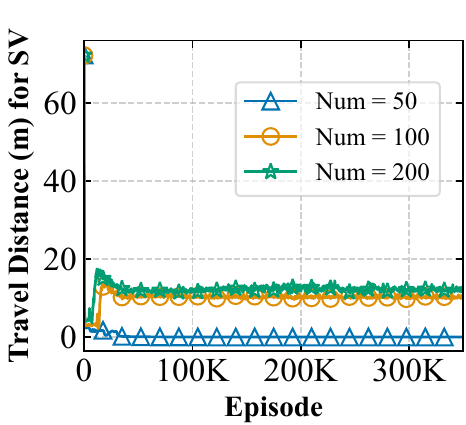}}\hfill
	\subfloat[]{\includegraphics[width=.19\linewidth]{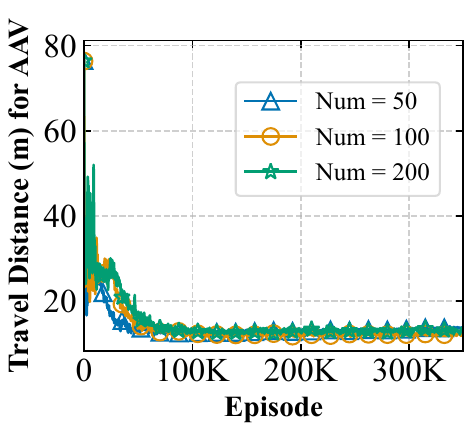}}\hfill
    \subfloat[]{\includegraphics[width=.19\linewidth]{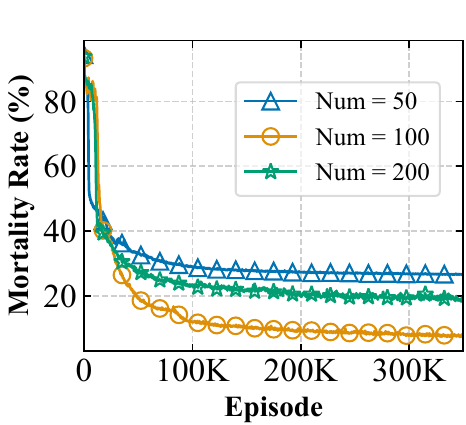}}\hfill
    \caption{Results of IHATRPO under different numbers of sensor nodes. (a) The charging efficiency for SV. (b) The charging efficiency for AAV. (c) The travel distance for SV. (d) The travel distance for AAV. (e) The mortality of sensor nodes ($100\times f_3 \%$).}
    \label{fig:sensornumR1}
\end{minipage}
\end{figure*}

\begin{figure*}[!t]
\begin{minipage}[t]{1\linewidth}
  \centering
	\subfloat[]{\includegraphics[width=.19\linewidth]{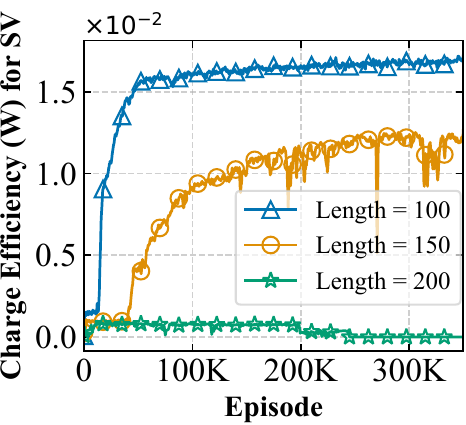}}\hfill
        \subfloat[]{\includegraphics[width=.19\linewidth]{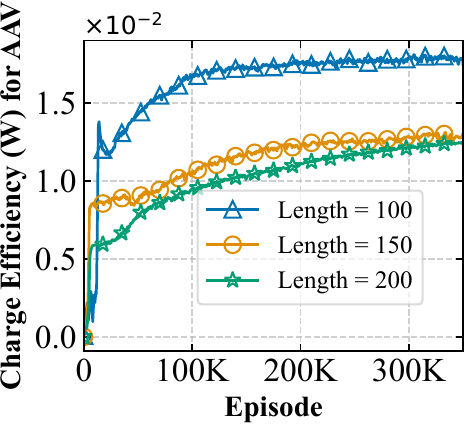}}\hfill
	\subfloat[]{\includegraphics[width=.19\linewidth]{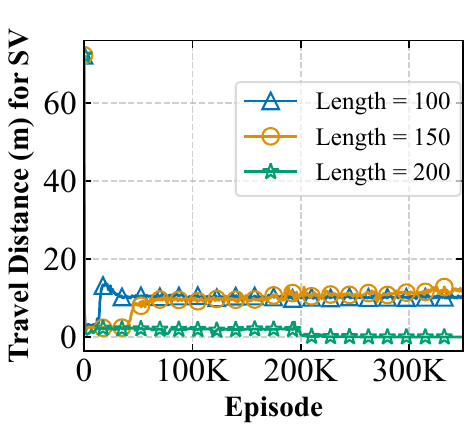}}\hfill
	\subfloat[]{\includegraphics[width=.19\linewidth]{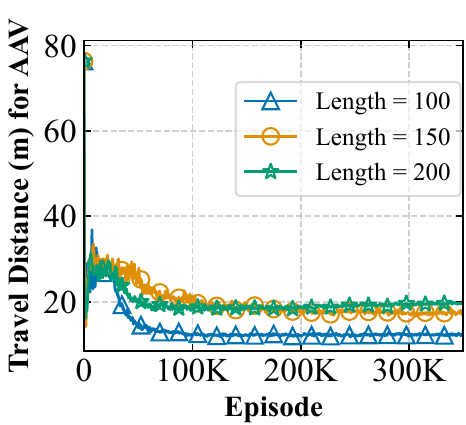}}\hfill
    \subfloat[]{\includegraphics[width=.19\linewidth]{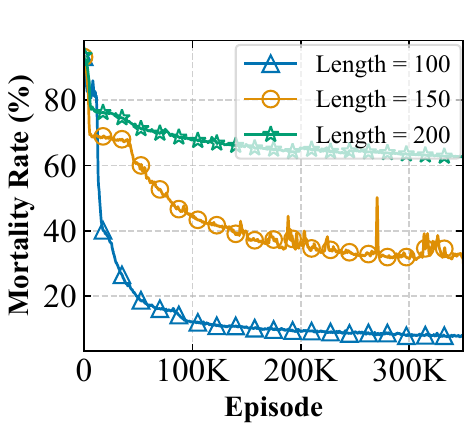}}\hfill
    \caption{Results of IHATRPO under different lengths of WRSN areas. (a) The charging efficiency for SV. (b) The charging efficiency for AAV. (c) The travel distance for SV. (d) The travel distance for AAV. (e) The mortality of sensor nodes ($100\times f_3 \%$).}
    \label{fig:arealengthR1}
\end{minipage}
\end{figure*}

\begin{figure*}[!t]
\begin{minipage}[t]{1\linewidth}
  \centering
	\subfloat[]{\includegraphics[width=.19\linewidth]{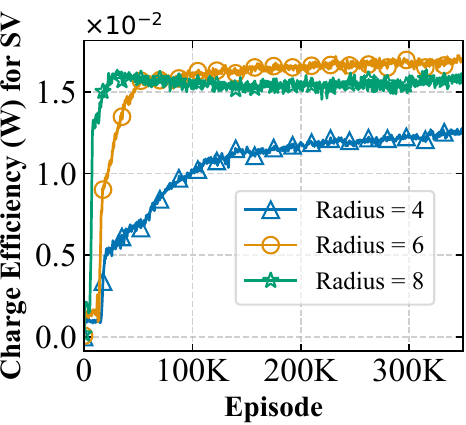}}\hfill
        \subfloat[]{\includegraphics[width=.19\linewidth]{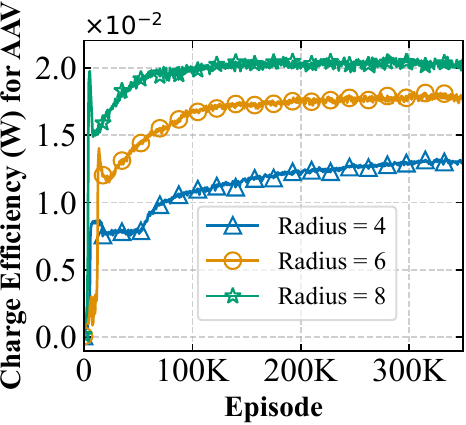}}\hfill
	\subfloat[]{\includegraphics[width=.19\linewidth]{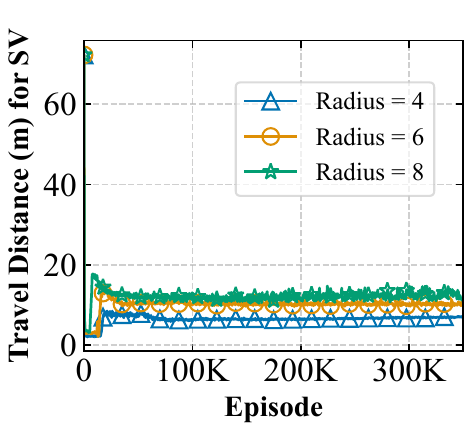}}\hfill
	\subfloat[]{\includegraphics[width=.19\linewidth]{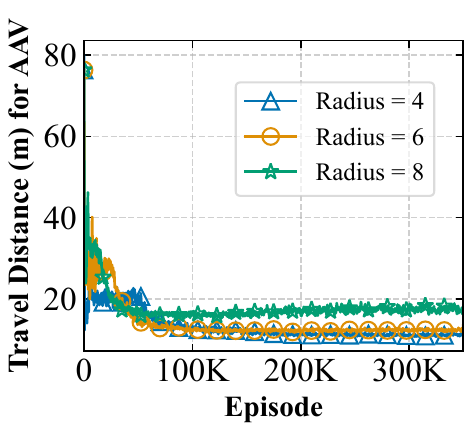}}\hfill
    \subfloat[]{\includegraphics[width=.19\linewidth]{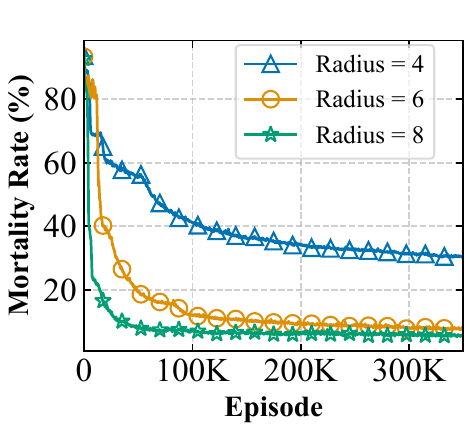}}\hfill
    \caption{Results of IHATRPO under different charging radii of mobile chargers AAV/SV. (a) The charging efficiency for SV. (b) The charging efficiency for AAV. (c) The travel distance for SV. (d) The travel distance for AAV. (e) The mortality of sensor nodes ($100\times f_3 \%$).}
    \label{fig:chargingradiusR1}
\end{minipage}
\end{figure*}

\begin{figure*}[!t]
\begin{minipage}[t]{1\linewidth}
  \centering
	\subfloat[]{\includegraphics[width=.19\linewidth]{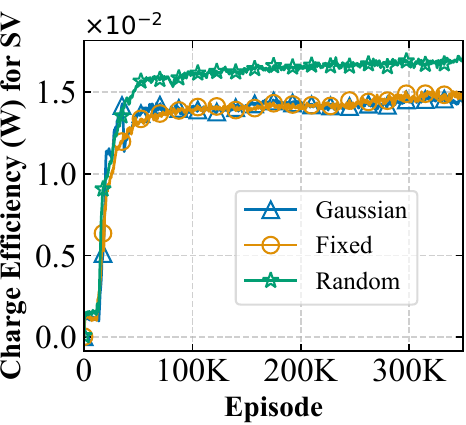}}\hfill
        \subfloat[]{\includegraphics[width=.19\linewidth]{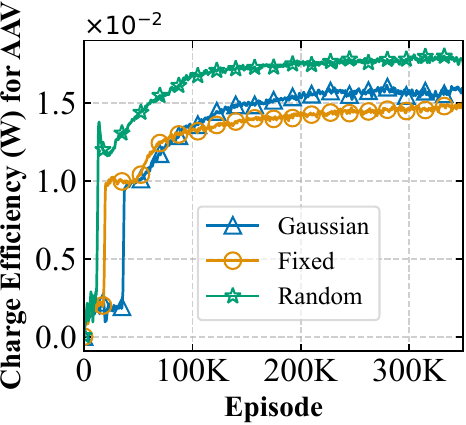}}\hfill
	\subfloat[]{\includegraphics[width=.19\linewidth]{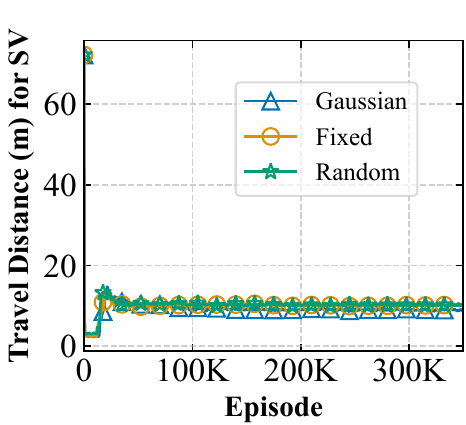}}\hfill
	\subfloat[]{\includegraphics[width=.19\linewidth]{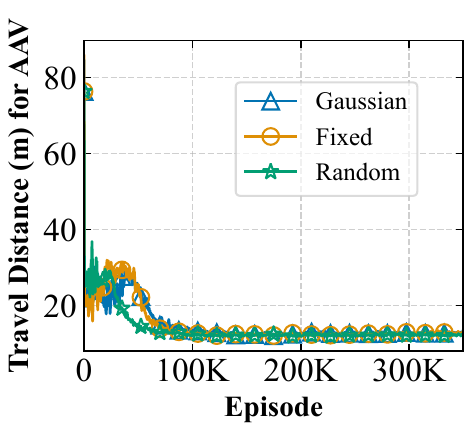}}\hfill
    \subfloat[]{\includegraphics[width=.19\linewidth]{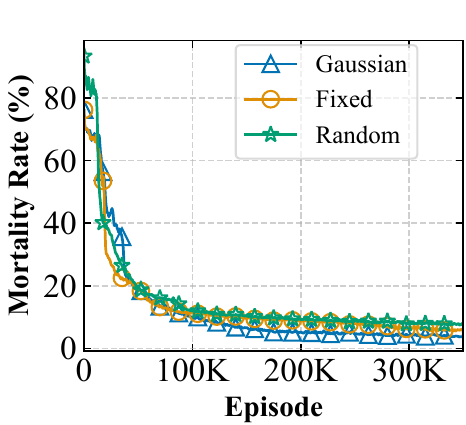}}\hfill
    \caption{Results of IHATRPO under different initial energy levels of sensor nodes. (a) The charging efficiency for SV. (b) The charging efficiency for AAV. (c) The travel distance for SV. (d) The travel distance for AAV. (e) The mortality of sensor nodes ($100\times f_3 \%$).}
    \label{fig:sensorenergyR1}
\end{minipage}
\end{figure*}

\begin{figure*}[!t]
\begin{minipage}[t]{1\linewidth}
  \centering
	\subfloat[]{\includegraphics[width=.19\linewidth]{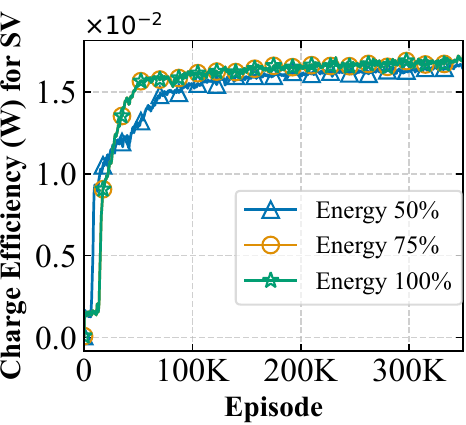}}\hfill
        \subfloat[]{\includegraphics[width=.19\linewidth]{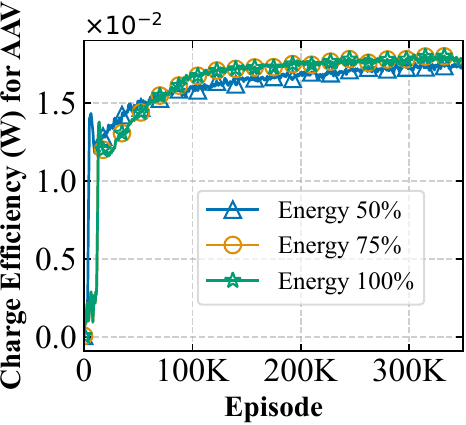}}\hfill
	\subfloat[]{\includegraphics[width=.19\linewidth]{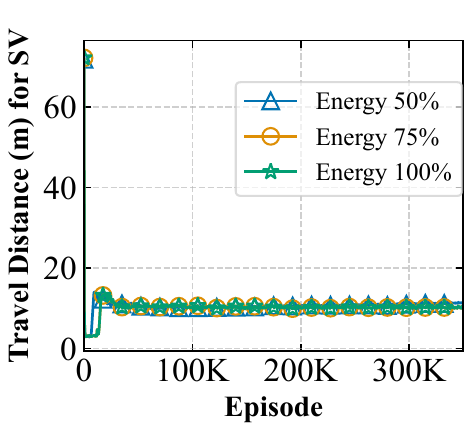}}\hfill
	\subfloat[]{\includegraphics[width=.19\linewidth]{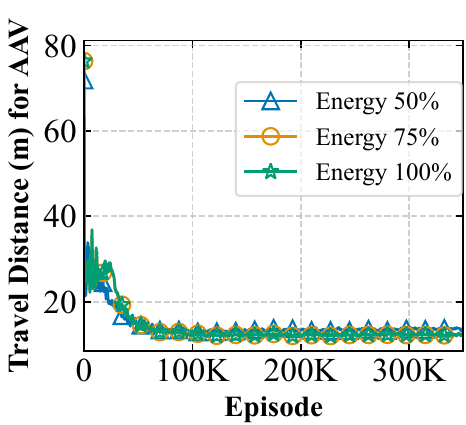}}\hfill
    \subfloat[]{\includegraphics[width=.19\linewidth]{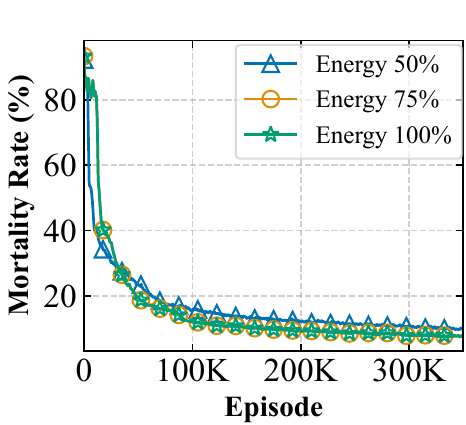}}\hfill
    \caption{Results of IHATRPO under different total energy budgets of AAV/SV. (a) The charging efficiency for SV. (b) The charging efficiency for AAV. (c) The travel distance for SV. (d) The travel distance for AAV. (e) The mortality of sensor nodes ($100\times f_3 \%$).}
    \label{fig:chargerenergyR1}
\end{minipage}
\end{figure*}

\begin{figure*}[!t]
\begin{minipage}[t]{1\linewidth}
  \centering
	\subfloat[]{\includegraphics[width=.19\linewidth]{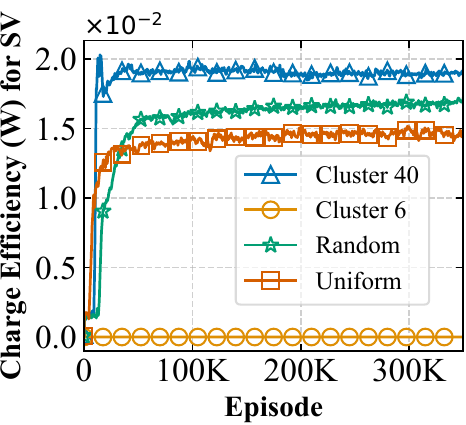}}\hfill
        \subfloat[]{\includegraphics[width=.19\linewidth]{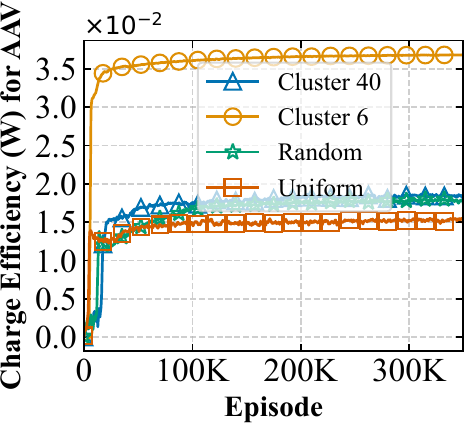}}\hfill
	\subfloat[]{\includegraphics[width=.19\linewidth]{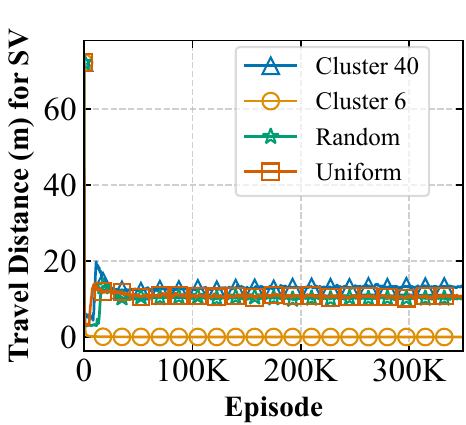}}\hfill
	\subfloat[]{\includegraphics[width=.19\linewidth]{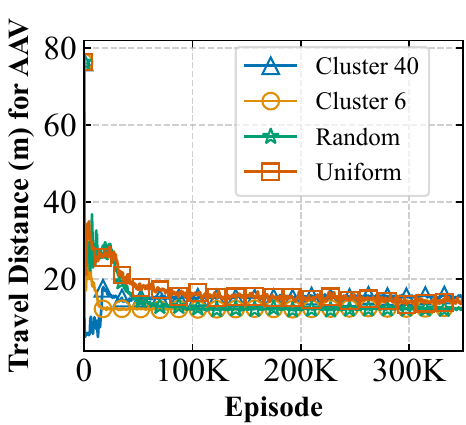}}\hfill
    \subfloat[]{\includegraphics[width=.19\linewidth]{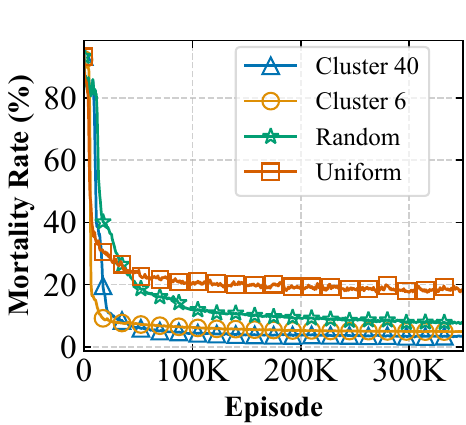}}\hfill
    \caption{Results of IHATRPO under different position distributions of sensor nodes. (a) The charging efficiency for SV. (b) The charging efficiency for AAV. (c) The travel distance for SV. (d) The travel distance for AAV. (e) The mortality of sensor nodes ($100\times f_3 \%$).}
    \label{fig:sensorpositionR1}
\end{minipage}
\end{figure*}

\begin{figure*}[!t]
\begin{minipage}[t]{1\linewidth}
  \centering
	\subfloat[]{\includegraphics[width=.19\linewidth]{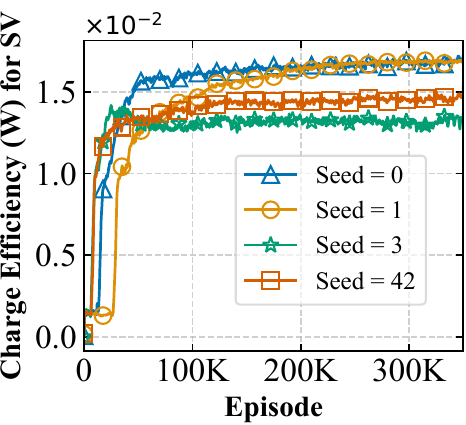}}\hfill
        \subfloat[]{\includegraphics[width=.19\linewidth]{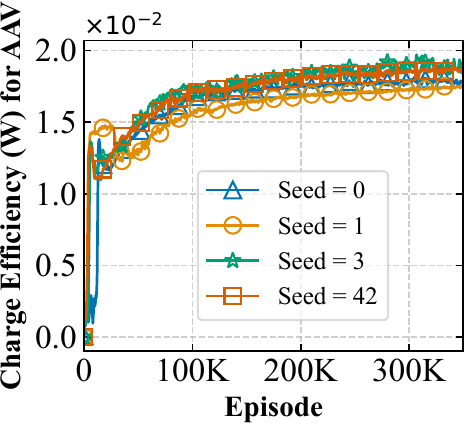}}\hfill
	\subfloat[]{\includegraphics[width=.19\linewidth]{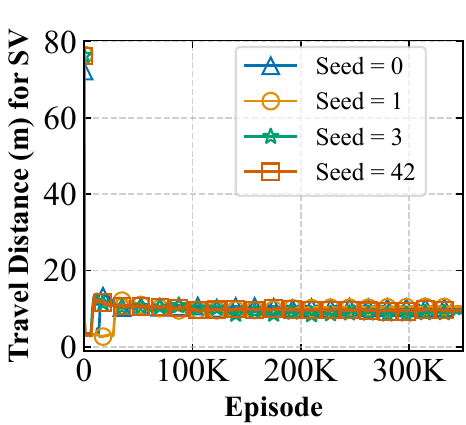}}\hfill
	\subfloat[]{\includegraphics[width=.19\linewidth]{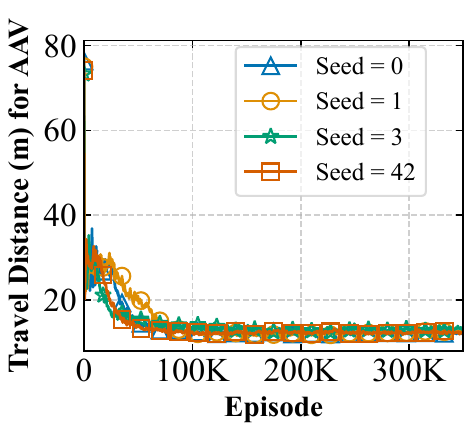}}\hfill
    \subfloat[]{\includegraphics[width=.19\linewidth]{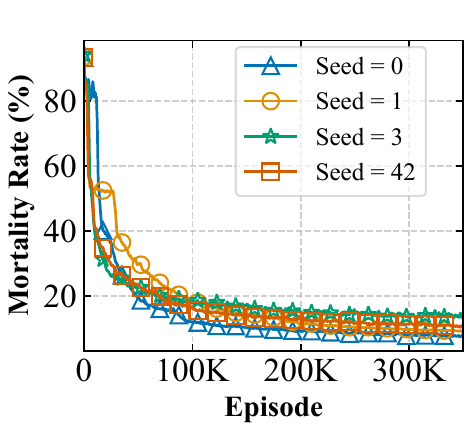}}\hfill
    \caption{Results of IHATRPO under different seeds. (a) The charging efficiency for SV. (b) The charging efficiency for AAV. (c) The travel distance for SV. (d) The travel distance for AAV. (e) The mortality of sensor nodes ($100\times f_3 \%$).}
    \label{fig:seedsR1}
\end{minipage}
\end{figure*}

\begin{figure}[htbp]
    \centering
    \includegraphics[width=0.7\linewidth]{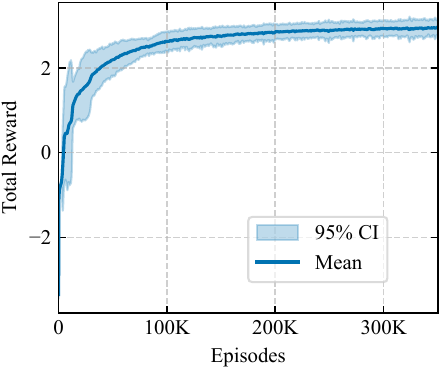}
    \caption{Results of IHATRPO averaged across all seeds, with the shaded region indicating the 95\% confidence interval (CI).}
    \label{fig:seedCI}
\end{figure}

\par To further evaluate the robustness of IHATRPO, we conduct sensitivity analyses from multiple dimensions, across the density of sensor nodes, settings of the AAV and SV, and the settings in energy and position of sensor nodes. The details are as follows:

\par \textit{Firstly}, we set the number of sensor nodes $N \in \{50, 100, 200\}$ and the side length of WRSN area $\{100, 150, 200\}$~m separately, thereby testing the impact of densities of sensor nodes. As shown in Figs.~\ref{fig:sensornumR1} and~\ref{fig:arealengthR1}, an appropriate sensor node density is beneficial for IHATRPO to reduce the mortality rate of sensor nodes. On the one hand, when the density is too high, although the charging efficiency of the AAV and SV improves, the AAV and SV face challenges in providing sufficient energy or timely charging services to all sensor nodes. On the other hand, when the density is too low, the AAV and SV are required to travel greater distances, thereby resulting in higher energy consumption for mobility while making it increasingly difficult to meet the charging demands of distant sensor nodes in time. However, IHATRPO achieves good optimization results in the mortality rate of sensor nodes across all settings.

\par \textit{Secondly}, we set the charging radius $\{4, 6, 8\}$~m and the energy budget $\{50\%, 75\%, 100\%\}$ of the default energy budget of the AAV and SV separately, thereby testing the impact of settings of the AAV and SV. As shown in Figs.~\ref{fig:chargingradiusR1} and~\ref{fig:chargerenergyR1}, a small charging radius and insufficient energy budget both degrade charging coverage and increase node mortality, whereas radius $6$~m and enough energy budget yield satisfactory optimization performance.

\par \textit{Thirdly}, we set the initial energy configuration, \textit{i.e.}, Gaussian distribution, uniform random distribution, and fixed $1.5$~J, and the position configuration, \textit{i.e.}, random, uniform, and clustered deployment scenarios of sensor nodes separately, thereby testing the impact of settings of sensor nodes. As shown in Fig.~\ref{fig:sensorenergyR1}, IHATRPO demonstrates robustness to energy heterogeneity across all settings. As for the position distribution of sensor nodes, as shown in Fig.~\ref{fig:sensorpositionR1}(c), when sensor nodes form only a small number of clusters within the WRSN area, the SV is unable to move between these clusters in time to respond to charging demands, and consequently tends to remain stationary, consistent with the observations in Fig.~\ref{fig:sensornumR1}(c) and Fig.~\ref{fig:arealengthR1}(c). However, IHATRPO achieves good optimization results in the mortality rate of sensor nodes across all position settings of sensor nodes.

\par \textit{Moreover}, we set different seeds $\{0, 1, 3, 42\}$ and report the 95\% confidence interval of the averaged total reward, as shown in Figs.~\ref{fig:seedsR1} and~\ref{fig:seedCI}, thereby confirming the strong reproducibility of IHATRPO.

% R1-4

\begin{figure}[htbp]
    \centering
    \includegraphics[width=0.8\linewidth]{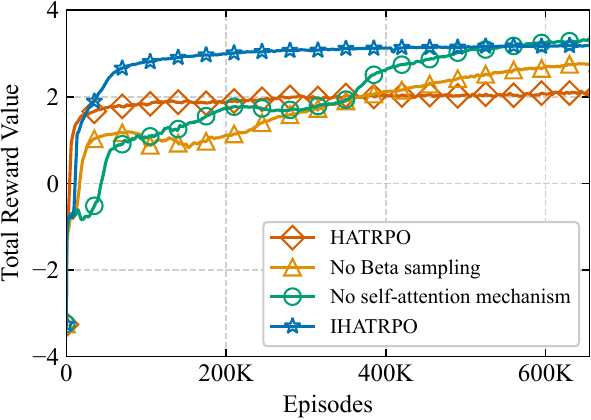}
    \caption{Effectiveness of different techniques (the self-attention mechanism and Beta sampling strategy).}
    \label{fig:ablation}
\end{figure}

\subsubsection{Ablation Analysis}
\par Fig.~\ref{fig:ablation} presents the contribution of each proposed component in IHATRPO. Specifically, we examine the effects of removing the self-attention mechanism and the Beta sampling strategy, respectively. As can be seen, the complete IHATRPO achieves the highest total reward value and demonstrates stable convergence, thereby highlighting the synergistic effect of its components. When the self-attention mechanism is removed, slower convergence suggests that the integration of the self-attention enhances the capability to extract and integrate critical state information from the complex environment. Similarly, the removal of the Beta sampling causes a noticeable decline in the final reward value, which indicates that the integration of the Beta sampling strategy helps the policy explore the bounded action space more effectively.
% R2-1

\par Quantitatively, the integration of the self-attention mechanism and Beta sampling strategy yields an overall reward value improvement of approximately 51\% compared with the original HATRPO algorithm. This result in Fig.~\ref{fig:ablation} confirms that both components contribute significantly to enhance the learning performance and overall reward of IHATRPO.
% R5-1

\begin{figure}[htbp]
    \centering
    \includegraphics[width=0.8\linewidth]{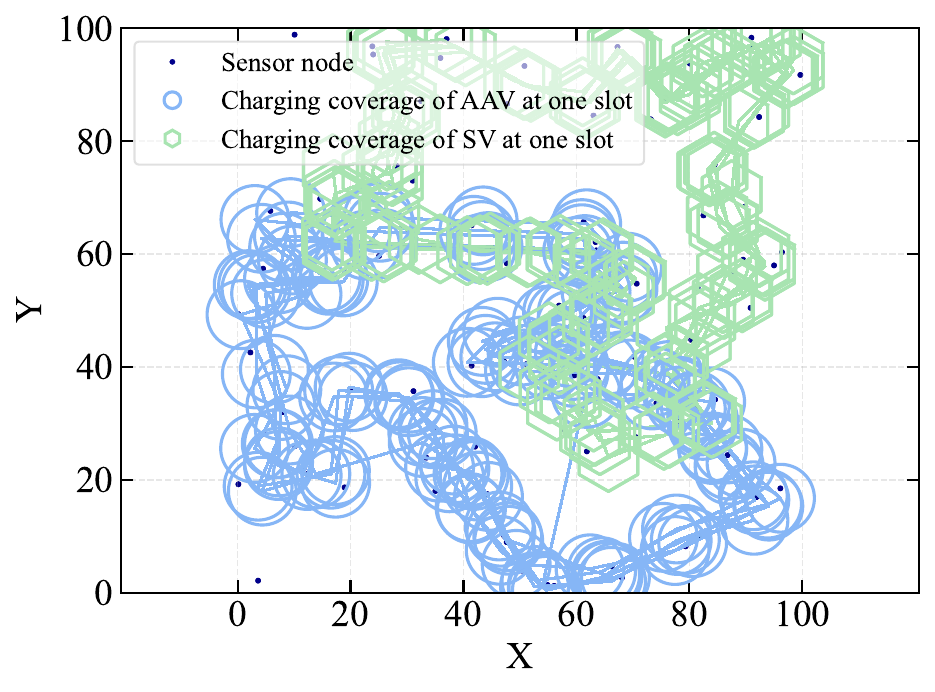}
    \caption{The trajectory of the AAV and SV obtained by IHATRPO.}
    \label{fig:trajectory_AAV_SV}
\end{figure}

\subsubsection{Spatial Movement Patterns Analysis}
\par Fig.~\ref{fig:trajectory_AAV_SV} shows the trajectory patterns and spatial distribution of the AAV and SV in the WRSN obtained through IHATRPO optimization. As observed in the trajectory visualization, the AAV primarily operates in the lower region of the sensor network, while the SV predominantly covers the upper region. The middle area demonstrates overlapping coverage where sensor nodes may be served by either agent, with the actual charging responsibility determined dynamically based on real-time charging requirements and spatial proximity.

\par This territorial division emerges naturally from the embedded coordination mechanism in IHATRPO. The self-attention mechanism enables each agent to dynamically assess charging priorities and spatial distribution based on the current WRSN conditions, which results in an efficient labor division that minimizes redundant coverage. Moreover, the Beta sampling enables agents to discover optimal territorial boundaries that balance workload distribution and service efficiency. This territorial coordination demonstrates the effectiveness of IHATRPO in achieving intelligent spatial resource allocation without explicit territorial assignment protocols or centralized coordination mechanisms.

%
%Conclusion
%
\section{Conclusion}
\label{sec:conclusion}

\par This paper has investigated a collaborative charging optimization problem for WRSNs using heterogeneous mobile chargers in heterogeneous deployment scenarios. Following this, we have formulated a multi-objective optimization problem to simultaneously maximize charging efficiency, minimize mobility energy consumption, and reduce sensor node mortality by coordinating the AAV and SV. The problem has proved highly challenging due to its dynamic nature with real-time adaptation requirements and complex trade-offs between competing objectives in heterogeneous multi-agent environments. To address these challenges, we have proposed the novel IHATRPO algorithm that incorporates the self-attention mechanism for enhanced environmental processing and the Beta sampling strategy for unbiased gradient computation in continuous action spaces. Simulation results have demonstrated that the proposed IHATRPO algorithm achieves faster convergence and superior performance compared to baselines, with sensor node mortality dramatically reduced from over 90\% to below 10\%. Spatial movement patterns analysis shows that the AAV and SV naturally develop complementary coverage patterns through the embedded coordination mechanism, with each agent specializing in different network regions to achieve efficient spatial division of labor. However, the current simulation settings are still based on an obstacle-free WRSN area. In more complex physical environments or larger-scale WRSNs, IHATRPO faces more challenges for the AAV and SV to provide timely charging services for all nodes. Therefore, future work may focus on more complex simulation environments, large-scale WRSNs, and coordination among more charging agents.

\section*{Acknowledgment}
\par The authors would like to thank Prof. Dusit Niyato for his valuable advisory support and insightful suggestions during the early stage of this work.

\normalem
\bibliographystyle{IEEEtran}
\bibliography{mybib}

@Article{Li2022,
  author  = {Jiahui Li and Geng Sun and Aimin Wang and Ming Lei and Shuang Liang and Hui Kang and Yanheng Liu},
  journal = {Comput. Netw.},
  title   = {A many-objective optimization charging scheme for wireless rechargeable sensor networks via mobile charging vehicles},
  year    = {2022},
  pages   = {109196},
  volume  = {215},
}

@Article{Qureshi2022,
  author  = {Bushra Qureshi and Sammah Abdel Aziz and Xingfu Wang and Ammar Hawbani and Saeed Hamood Alsamhi and Taiyaba Qureshi and Abdulbary Naji},
  journal = {Wirel. Netw.},
  title   = {A state-of-the-art survey on wireless rechargeable sensor networks: {P}erspectives and challenges},
  year    = {2022},
  number  = {7},
  pages   = {3019--3043},
  volume  = {28},
}

@Article{Zhao2024,
  author    = {Zirui Zhao and Tingdan Deng and Yixuan Liu and Feng Lin},
  journal = {Int. J. Distrib. Sens. Netw.},
  title     = {Charging between {PAD}s: {P}eriodic charging scheduling in the {UAV}-based {WRSN} with {PAD}s},
  number = {1},
  pages = {8851835},
  year      = {2024},
  volume    = {2024}
}

@Article{Kaswan2022,
  author  = {Amar Kaswan and Prasanta K. Jana and Sajal K. Das},
  journal = {{IEEE} Commun. Surv. Tutorials},
  title   = {A survey on mobile charging techniques in wireless rechargeable sensor networks},
  year    = {2022},
  number  = {3},
  pages   = {1750--1779},
  volume  = {24},
}

@InProceedings{Lin2020,
  author    = {Chi Lin and Feng Gao and Haipeng Dai and Jiankang Ren and Lei Wang and Guowei Wu},
  booktitle={Proc. {IEEE} {INFOCOM}},
  title     = {Maximizing charging utility with obstacles through {Fresnel} diffraction model},
  year      = {2020},
  pages     = {2046--2055},
}

@Article{Dai2020,
  author  = {Haipeng Dai and Qiufang Ma and Xiaobing Wu and Guihai Chen and David K. Y. Yau and Shaojie Tang and Xiang{-}Yang Li and Chen Tian},
  journal = {{IEEE} Trans. Mob. Comput.},
  title   = {{CHASE}: {C}harging and scheduling scheme for stochastic event capture in wireless rechargeable sensor networks},
  year    = {2020},
  number  = {1},
  pages   = {44--59},
  volume  = {19},
}

@Article{Yang2024,
  author  = {Yang Yang and Xuxun Liu and Kun Tang and Wenquan Che and Quan Xue},
  journal = {{IEEE} Trans. Sustain. Comput.},
  title   = {Multi-type charging scheduling based on area requirement difference for wireless rechargeable sensor networks},
  year    = {2024},
  number  = {2},
  pages   = {182--196},
  volume  = {9},
}

@Article{Lee2022,
  author  = {Donghyun Lee and Cheol Lee and Gunhee Jang and Woongsoo Na and Sungrae Cho},
  journal = {{IEEE} Internet Things J.},
  title   = {Energy-efficient directional charging strategy for wireless rechargeable sensor networks},
  year    = {2022},
  number  = {19},
  pages   = {19034--19048},
  volume  = {9},
}

@Article{Zhang2024,
  author  = {Xiuling Zhang and Riheng Jia and Quanjun Yin and Zhonglong Zheng and Minglu Li},
  journal = {{IEEE} Trans. Mob. Comput.},
  title   = {Intelligent trajectory design and charging scheduling in wireless rechargeable sensor networks with obstacles},
  year    = {2024},
  number  = {9},
  pages   = {8664--8679},
  volume  = {23},
}

@Article{Liu2022,
  author  = {Yanheng Liu and Hongyang Pan and Geng Sun and Aimin Wang and Jiahui Li and Shuang Liang},
  journal = {{IEEE} Internet Things J.},
  title   = {Joint scheduling and trajectory optimization of charging {UAV} in wireless rechargeable sensor networks},
  year    = {2022},
  number  = {14},
  pages   = {11796--11813},
  volume  = {9},
}

@Article{Liang2021,
  author  = {Shuang Liang and Zhiyi Fang and Geng Sun and Chi Lin and Jiahui Li and Songyang Li and Aimin Wang},
  journal = {Comput. Netw.},
  title   = {Charging {UAV} deployment for improving charging performance of wireless rechargeable sensor networks via joint optimization approach},
  year    = {2021},
  pages   = {108573},
  volume  = {201},
}

@Article{Lin2023a,
  author  = {Chi Lin and Shibo Hao and Wei Yang and Pengfei Wang and Lei Wang and Guowei Wu and Qiang Zhang},
  journal = {{IEEE/ACM} Trans. Netw.},
  title   = {Maximizing energy efficiency of period-area coverage with a {UAV} for wireless rechargeable sensor networks},
  year    = {2023},
  number  = {4},
  pages   = {1657--1673},
  volume  = {31},
}

@ARTICLE{Ning2024,
  author={Liu, Ning and Zhang, Jian and Luo, Chuanwen and Cao, Jia and Hong, Yi and Chen, Zhibo and Chen, Ting},
  journal = {{IEEE} Internet Things J.}, 
  title={Dynamic charging strategy optimization for {UAV}-assisted wireless rechargeable sensor networks based on deep {Q}-network},
  year={2024},
  volume={11},
  number={12},
  pages={21125--21134}}

@ARTICLE{Tao2020,
  author={Wu, Tao and Yang, Panlong and Dai, Haipeng and Xiang, Chaocan and Rao, Xunpeng and Huang, Jun and Ma, Tao},
  journal = {{IEEE} Internet Things J.}, 
  title={Joint sensor selection and energy allocation for tasks-driven mobile charging in wireless rechargeable sensor networks},
  year={2020},
  volume={7},
  number={12},
  pages={11505--11523}
}

@InProceedings{Schulman2015,
  author    = {John Schulman and Sergey Levine and Pieter Abbeel and Michael I. Jordan and Philipp Moritz},
  booktitle = {Proc. {ICML}},
  title     = {Trust region policy optimization},
  year      = {2015},
  pages     = {1889--1897},
  volume    = {37}
}

@Article{Jiang2024,
  author  = {Chengpeng Jiang and Wencong Chen and Xingcan Chen and Sen Zhang and Wendong Xiao},
  journal = {Expert Syst. Appl.},
  title   = {Deep reinforcement learning approach with hybrid action space for mobile charging in wireless rechargeable sensor networks},
  year    = {2024},
  pages   = {123752},
  volume  = {249},
}

@InProceedings{Liang2022,
  author    = {Yongheng Liang and Hejun Wu and Haitao Wang},
  booktitle = {Proc. {AAMAS}},
  title     = {{ASM-PPO}: asynchronous and scalable multi-agent {PPO} for cooperative charging},
  year      = {2022},
  pages     = {798--806},
}

@Article{Liu2022a,
  author  = {Ning Liu and Chuanwen Luo and Jia Cao and Yi Hong and Zhibo Chen},
  journal = {Sensors},
  title   = {Trajectory optimization of laser-charged {UAV}s for charging wireless rechargeable sensor networks},
  year    = {2022},
  number  = {23},
  pages   = {9215},
  volume  = {22},
}

@Article{Ning2025,
  author  = {Zhaolong Ning and Hongjing Ji and Xiaojie Wang and Edith C. H. Ngai and Lei Guo and Jiangchuan Liu},
  journal = {{IEEE} Trans. Mob. Comput.},
  title   = {Joint optimization of data acquisition and trajectory planning for {UAV}-assisted wireless powered internet of things},
  year    = {2025},
  number  = {2},
  pages   = {1016--1030},
  volume  = {24},
}

@Article{Chen2022,
  author  = {Tzung{-}Shi Chen and Jen{-}Jee Chen and Xiang{-}You Gao and Tzung{-}Cheng Chen},
  journal = {Sensors},
  title   = {Mobile charging strategy for wireless rechargeable sensor networks},
  year    = {2022},
  number  = {1},
  pages   = {359},
  volume  = {22},
}

@Article{Xie2012,
  author  = {Liguang Xie and Yi Shi and Y. Thomas Hou and Hanif D. Sherali},
  journal = {{IEEE/ACM} Trans. Netw.},
  title   = {Making sensor networks immortal: {A}n energy-renewal approach with wireless power transfer},
  year    = {2012},
  number  = {6},
  pages   = {1748--1761},
  volume  = {20},
}

@Article{Yi2019,
  author  = {Junmin Yi and Ikjune Yoon},
  journal = {Sensors},
  title   = {Efficient energy supply using mobile charger for solar-powered wireless sensor networks},
  year    = {2019},
  number  = {12},
  pages   = {2679},
  volume  = {19},
}

@Article{Shu2016,
  author    = {Yuanchao Shu and Hamed Yousefi and Peng Cheng and Jiming Chen and Yu Jason Gu and Tian He and Kang G. Shin},
  journal   = {{IEEE} Trans. Mob. Comput.},
  title     = {Near-optimal velocity control for mobile charging in wireless rechargeable sensor networks},
  year      = {2016},
  number    = {7},
  pages     = {1699--1713},
  volume    = {15}
}

@Article{Lin2016,
  author    = {Chi Lin and Zhiyuan Wang and Ding Han and Youkun Wu and Chang{-}Wu Yu and Guowei Wu},
  journal   = {J. Syst. Archit.},
  title     = {{TADP:} Enabling temporal and distantial priority scheduling for on-demand charging architecture in wireless rechargeable sensor networks},
  year      = {2016},
  pages     = {26--38},
  volume    = {70}
}

@Article{He2013,
  author  = {Shibo He and Jiming Chen and Fachang Jiang and David K. Y. Yau and Guoliang Xing and Youxian Sun},
  journal = {{IEEE} Trans. Mob. Comput.},
  title   = {Energy provisioning in wireless rechargeable sensor networks},
  year    = {2013},
  number  = {10},
  pages   = {1931--1942},
  volume  = {12},
}

@Article{Zeng2019,
  author  = {Yong Zeng and Jie Xu and Rui Zhang},
  journal = {{IEEE} Trans. Wirel. Commun.},
  title   = {Energy minimization for wireless communication with rotary-wing {UAV}},
  year    = {2019},
  number  = {4},
  pages   = {2329--2345},
  volume  = {18},
}

@InProceedings{Mei2004,
  author    = {Yongguo Mei and Yung{-}Hsiang Lu and Y. Charlie Hu and C. S. George Lee},
  booktitle = {Proc. {IEEE} {ICRA}},
  title     = {Energy-efficient motion planning for mobile robots},
  year      = {2004},
  pages     = {4344--4349},
}

@Article{Fu2016,
  author  = {Lingkun Fu and Peng Cheng and Yu Gu and Jiming Chen and Tian He},
  journal = {{IEEE} Trans. Veh. Technol.},
  title   = {Optimal charging in wireless rechargeable sensor networks},
  year    = {2016},
  number  = {1},
  pages   = {278--291},
  volume  = {65},
}

@Article{Yu2021,
  author    = {Yu Yu and Jie Tang and Jiayi Huang and Xiuyin Zhang and Daniel Ka Chun So and Kai{-}Kit Wong},
  journal   = {{IEEE} Trans. Commun.},
  title     = {Multi-objective optimization for {UAV}-assisted wireless powered {IoT} networks based on extended {DDPG} algorithm},
  year      = {2021},
  number    = {9},
  pages     = {6361--6374},
  volume    = {69}
}

@Article{Zhang2022,
  author    = {Liang Zhang and Abdulkadir Celik and Shuping Dang and Basem Shihada},
  journal   = {{IEEE} Trans. Mob. Comput.},
  title     = {Energy-efficient trajectory optimization for {UAV}-assisted {IoT} networks},
  year      = {2022},
  number    = {12},
  pages     = {4323--4337},
  volume    = {21}
}

@Article{Lyu2024,
  author    = {Ting Lyu and Jianwei An and Meng Li and Feifei Liu and Haitao Xu},
  journal   = {Computing},
  title     = {{UAV}-assisted wireless charging and data processing of power {IoT} devices},
  year      = {2024},
  number    = {3},
  pages     = {789--819},
  volume    = {106}
}

@Article{Hou2024,
  author  = {Chen Hou and Qilong Huang},
  journal = {{IEEE} Trans. Autom. Sci. Eng.},
  title   = {Energy supply control of wireless powered piecewise linear neural network},
  year    = {2024},
  number  = {4},
  pages   = {6892--6907},
  volume  = {21},
}

@Article{Schulman2017,
  author        = {John Schulman and Filip Wolski and Prafulla Dhariwal and Alec Radford and Oleg Klimov},
  journal       = {CoRR},
  title         = {Proximal policy optimization algorithms},
  year          = {2017},
  volume        = {abs/1707.06347},
}

@InProceedings{Lillicrap2016,
  author    = {Timothy P. Lillicrap and Jonathan J. Hunt and Alexander Pritzel and Nicolas Heess and Tom Erez and Yuval Tassa and David Silver and Daan Wierstra},
  booktitle = {Proc. {ICLR}},
  title     = {Continuous control with deep reinforcement learning},
  year      = {2016},
}

@InProceedings{Lowe2017,
  author    = {Ryan Lowe and Yi Wu and Aviv Tamar and Jean Harb and Pieter Abbeel and Igor Mordatch},
  booktitle = {Adv. Neural Inf. Process. Syst.},
  title     = {Multi-agent actor-critic for mixed cooperative-competitive environments},
  year      = {2017},
  pages     = {6379--6390},
}

@Article{Akyildiz2002,
  author  = {Akyildiz, I.F. and Weilian Su and Sankarasubramaniam, Y. and Cayirci, E.},
  journal = {{IEEE} Commun. Mag.},
  title   = {A survey on sensor networks},
  year    = {2002},
  number  = {8},
  pages   = {102--114},
  volume  = {40},
}

@Article{Kandris2020,
  author  = {Kandris, Dionisis and Nakas, Christos and Vomvas, Dimitrios and Koulouras, Grigorios},
  journal = {Appl. Syst. Innov.},
  title   = {Applications of wireless sensor networks: an up-to-date survey},
  year    = {2020},
  number  = {1},
  volume  = {3},
}

@Article{Dhabliya2022,
  author  = {Dhabliya, Dharmesh and Soundararajan, Rajasoundaran and Selvarasu, Parthiban and Balasubramaniam, Maruthi Shankar and Rajawat, Anand Singh and Goyal, S. B. and Raboaca, Maria Simona and Mihaltan, Traian Candin and Verma, Chaman and Suciu, George},
  journal = {Energies},
  title   = {Energy-efficient network protocols and resilient data transmission schemes for wireless sensor networks---an experimental survey},
  year    = {2022},
  number  = {23},
  volume  = {15},
}

@Article{Khan2024,
  author  = {Khan, Md Yakub Ali and Hussain, Md and Halim, Md and Ibrahim, Salah and Haque, Abrarul},
  journal = {Control Syst. Optim. Lett.},
  title   = {A comprehensive review on techniques and challenges of energy harvesting from distributed renewable energy sources for wireless sensor networks},
  year    = {2024},
  pages   = {15--22},
  volume  = {2},
}

@Article{LeonAvila2025,
  author  = {Bernardo Yaser {León Ávila} and Carlos Alberto {García Vázquez} and Osmel {Pérez Baluja} and Daniel Tudor Cotfas and Petru Adrian Cotfas},
  journal = {Energy Strategy Rev.},
  title   = {Energy harvesting techniques for wireless sensor networks: {A} systematic literature review},
  year    = {2025},
  pages   = {101617},
  volume  = {57},
}

@Article{Mou2023,
  author  = {Mou, Xiaolin and Gladwin, Daniel and Jiang, Jing and Li, Kang and Yang, Zhile},
  journal = {{IEEE} J. Emerg. Sel. Topics Ind. Electron.},
  title   = {Near-field wireless power transfer technology for unmanned aerial vehicles: a systematical review},
  year    = {2023},
  number  = {1},
  pages   = {147--158},
  volume  = {4},
}

@Article{Qaisar2024,
  author    = {Muhammad Umar Farooq Qaisar and Weijie Yuan and Paolo Bellavista and Fan Liu and Guangjie Han and Rabiu Sale Zakariyya and Adeel Ahmed},
  journal   = {{IEEE} Trans. Mob. Comput.},
  title     = {Poised: probabilistic on-demand charging scheduling for {ISAC}-assisted {WRSN}s with multiple mobile charging vehicles},
  year      = {2024},
  number    = {12},
  pages     = {10818--10834},
  volume    = {23}
}

@Article{Qaisar2024a,
  author    = {Muhammad Umar Farooq Qaisar and Weijie Yuan and Paolo Bellavista and Guangjie Han and Adeel Ahmed},
  journal   = {{IEEE} Internet Things Mag.},
  title     = {{ISAC}-assisted wireless rechargeable sensor networks with multiple mobile charging vehicles},
  year      = {2024},
  number    = {6},
  pages     = {80--86},
  volume    = {7}
}

@Article{Li2024,
  author  = {Li, Lizhi and Feng, Yong and Liu, Nianbo and Li, Yingna and Zhang, Jing},
  journal = {{IEEE} Sensors J.},
  title   = {Deep reinforcement learning-based dynamic charging--recycling scheme for wireless rechargeable sensor networks},
  year    = {2024},
  number  = {9},
  pages   = {15457--15471},
  volume  = {24},
}

@Article{Orumwense2022,
  author  = {Orumwense, Efe Francis and Abo-Al-Ez, Khaled},
  journal = {Energies},
  title   = {On increasing the energy efficiency of wireless rechargeable sensor networks for cyber-physical systems},
  year    = {2022},
  number  = {3},
  volume  = {15},
}

@InProceedings{Kuba2022,
  author    = {Jakub Grudzien Kuba and Ruiqing Chen and Muning Wen and Ying Wen and Fanglei Sun and Jun Wang and Yaodong Yang},
  booktitle = {Proc. {ICLR}},
  title     = {Trust region policy optimisation in multi-agent reinforcement learning},
  year      = {2022},
}

@inproceedings{Vaswani2017,
 author = {Vaswani, Ashish and Shazeer, Noam and Parmar, Niki and Uszkoreit, Jakob and Jones, Llion and Gomez, Aidan N and Kaiser, \L ukasz and Polosukhin, Illia},
 booktitle = {Adv. Neural Inf. Process. Syst.},
 title = {Attention is all you need},
 volume = {30},
 year = {2017}
}

@Article{Gronauer2022,
  author    = {Sven Gronauer and Klaus Diepold},
  journal   = {Artif. Intell. Rev.},
  title     = {Multi-agent deep reinforcement learning: {A} survey},
  year      = {2022},
  number    = {2},
  pages     = {895--943},
  volume    = {55}
}

@Article{Sun2026,
  author    = {Geng Sun and Likun Zhang and Jiahui Li and Jing Wu and Jiacheng Wang and Zemin Sun and Changyuan Zhao and Victor C. M. Leung},
  journal   = {{IEEE} Trans. Netw. Sci. Eng.},
  title     = {Age of information optimization in laser-charged {UAV}-assisted {IoT} networks: {A} multi-agent deep reinforcement learning method},
  year      = {2026},
  pages     = {1436--1457},
  volume    = {13},
}

@mastersthesis{Chou2017,
author = {Po-Wei Chou},
title = {The {B}eta policy for continuous control reinforcement learning},
year = {2017},
school = {Carnegie Mellon University},
number = {CMU-RI-TR-17-38},
}

@article{Greenwood2019,
author = {William W. Greenwood  and Jerome P. Lynch  and Dimitrios Zekkos },
title = {Applications of {UAV}s in civil infrastructure},
journal = {J. Infrastruct. Syst.},
volume = {25},
number = {2},
pages = {04019002},
year = {2019}
}

\end{document}